\documentclass{emulateapj}
\usepackage{apjfonts}
\journalinfo{The Astrophysical Journal, 706, 2009 December 1, in press}

\shorttitle{T TAURI ACCRETION-DRIVEN CORONAL HEATING}
\shortauthors{CRANMER}

\begin{document}

\title{Testing Models of Accretion-driven Coronal Heating and
Stellar Wind Acceleration for T Tauri Stars}

\author{Steven R. Cranmer}
\affil{Harvard-Smithsonian Center for Astrophysics,
60 Garden Street, Cambridge, MA 02138}
\email{scranmer@cfa.harvard.edu}

\begin{abstract}
Classical T Tauri stars are pre-main-sequence objects that undergo
simultaneous accretion, wind outflow, and coronal X-ray emission.
The impact of plasma on the stellar surface from magnetospheric
accretion streams is likely to be a dominant source of energy
and momentum in the upper atmospheres of these stars.
This paper presents a set of models for the dynamics and heating
of three distinct regions on T Tauri stars that are affected by
accretion:  (1) the shocked plasmas directly beneath the
magnetospheric accretion streams, (2) stellar winds that are
accelerated along open magnetic flux tubes, and (3) closed
magnetic loops that resemble the Sun's coronal active regions.
For the loops, a self-consistent model of coronal heating was
derived from numerical simulations of solar field-line tangling
and turbulent dissipation.
Individual models are constructed for the properties of 14
well-observed stars in the Taurus-Auriga star-forming region.
Predictions for the wind mass loss rates are, on average,
slightly lower than the observations, which suggests that disk
winds or X-winds may also contribute to the measured outflows.
For some of the stars, however, the modeled stellar winds do
appear to contribute significantly to the measured mass fluxes.
Predictions for X-ray luminosities from the shocks and loops
are in general agreement with existing observations.
The stars with the highest accretion rates tend to have
X-ray luminosities dominated by the high-temperature
(5--10 MK) loops.
The X-ray luminosities for the stars having lower accretion
rates are dominated by the cooler accretion shocks.
\end{abstract}


\section{Introduction}

Nearly all low-mass stars exhibit magnetic activity and some kind
of mass outflow.
Young stars add to this already complex situation by also undergoing
{\em accretion} along a subset of the magnetic field lines that
connect the star and its surrounding disk.
Out of the many unanswered questions regarding star formation and
early stellar evolution, many of them share the need to better
understand coronal heating and the three-dimensional
magnetohydrodynamic (MHD) interplay between winds and accretion.
These physical processes are key to determining how rapidly the
stars rotate, how active the stars appear from radio to X-ray
wavelengths, and how the stars affect nearby planets.

For classical T Tauri stars (CTTS), there are several possible
explanations of how and where outflows arise, including extended
disk winds, compact X-winds, impulsive eruptions, and stellar
winds that bypass the accretion disk altogether (see, e.g.,
Paatz \& Camenzind 1996; Ferreira et al.\  2006;
Shu et al.\  2007; Casse et al.\  2007;
G\'{o}mez de Castro \& Verdugo 2007; Edwards 2009;
Aarnio et al.\  2009; Romanova et al.\  2009;
Krasnopolsky et al.\  2009).
It is of definite interest to determine how much of the observed
mass loss can be explained solely with stellar winds, since
outflows that are ``locked'' to the surface are capable of removing
excess angular momentum from the star (Matt \& Pudritz 2005,
2007, 2008).
In addition, there appear be multiple source regions of ultraviolet
(UV) and X-ray activity, such as flares, accretion shocks on the
stellar surface, and closed loops that may resemble the Sun's
coronal active regions (e.g., Feigelson \& Montmerle 1999;
Kastner et al.\  2002; Stassun et al.\  2006, 2007;
Argiroffi et al.\  2007; G\"{u}del et al.\  2007b; Jardine 2008).

Much of the research devoted to improving our understanding of
coronal heating and wind acceleration has been focused on the Sun.
After many decades of high-resolution observations, analytic
studies, and numerical simulations, it seems increasingly clear
that closed magnetic loops in the solar corona are heated by
intermittent reconnection events that are driven by the
convective stressing of the loop footpoints (e.g.,
Aschwanden 2006; Klimchuk 2006).
However, the open field lines that connect the Sun to
interplanetary space are most likely energized by the
dissipation of waves and turbulent motions that originate at the
surface and then propagate outwards (Tu \& Marsch 1995;
Cranmer 2002; Suzuki 2006; Kohl et al.\  2006).
Self-consistent models of turbulence-driven coronal heating
and solar wind acceleration have begun to succeed in reproducing
a wide range of observations without the need for ad~hoc free
parameters (Rappazzo et al.\  2007, 2008; Cranmer et al.\  2007;
Cranmer 2009).
Such progress on the solar front provides a fruitful opportunity
to better understand the fundamental physics of coronal heating,
accretion, and wind acceleration in young stars.

The goal of this paper is to continue the construction and testing
of self-consistent models for the atmospheric heating and wind
acceleration of CTTS.
The starting point for this work is an existing methodology
developed by Cranmer (2008) for computing the properties of
T Tauri stellar winds.
These models incorporated a new physical mechanism for how the
mid-latitude accretion streams can influence the polar outflow.
The inhomogeneous impact of accreting material was suggested to
generate strong MHD turbulence and waves that propagate over the
stellar surface to energize the launching points of the stellar
wind streams.
However, these models were created only for an idealized
evolutionary sequence of a one solar mass ($M_{\ast} = 1 \,
M_{\odot}$) star, and they dealt only with the polar wind.
Both the mass loss rates and the X-ray fluxes predicted by
Cranmer (2008) were lower than typical observed values for
T Tauri stars by at least an order of magnitude.
It is clear that these models need to be adapted to a more
``real world'' situation that encompasses both (1) actual measured
parameters of individual T Tauri stars, and (2) a more
comprehensive treatment of both the open and closed magnetic
field structures on the stellar surface.

Section 2 discusses the specific CTTS stars selected for this
study and summarizes some of the most salient fundamental
properties derived from observations.
Section 3 outlines the methodology used to estimate the magnetic
field geometries, the accretion stream dynamics, and the
properties of impact-generated waves for each modeled stellar case.
The next three sections describe how the plasma properties in
three main accretion-driven surface regions are computed.
Section 4 deals with the acceleration of the stellar wind from
the polar regions.
Section 5 deals with the localized ``hot spot'' heating at the
accretion shocks.
Section 6 deals with the distributed heating in low-latitude
coronal loops.
Then Section 7 compares the modeled X-ray luminosities from the
accretion shocks and coronal loops to existing observations.
Finally, Section 8 contains a brief summary of the major
results and discusses how the process of testing and refining
the theoretical models should be improved in future work.

\section{Stellar Sample: Taurus-Auriga Molecular Cloud}

The main purpose of this paper is to test a range of proposed
physical processes in T Tauri stellar atmospheres by creating
models of specific well-observed stars.
The stars in the sample must be chosen on the basis of having
as many measurements of relevant quantities as possible to
constrain the models.
The Taurus-Auriga star-forming region was chosen because of its
relative proximity, its large number of low-mass
pre-main-sequence stars, and its high level of X-ray activity
(see, e.g., G\"{u}del et al.\  2007a).
These stars have not only multiple measurements of their mass
accretion rates (Hartigan et al.\  1995;
Hartmann et al.\  1998; Gullbring et al.\  1998;
White \& Ghez 2001), but also empirical determinations of the
wind mass loss rates, accretion-spot filling factors, and
magnetic field strengths.

\begin{deluxetable*}{ccccccccc}
\tablecaption{Measured Stellar Properties}
\tablewidth{0pt}

\tablehead{
\colhead{} &
\colhead{log (age)\tablenotemark{a}} &
\colhead{} &
\colhead{} &
\colhead{} &
\colhead{$B_{\ast}$} &
\colhead{log $\dot{M}_{\rm acc}$\tablenotemark{a}} &
\colhead{log $\dot{M}_{\rm wind}$} &
\colhead{} \\
\colhead{Object} &
\colhead{(yr)} &
\colhead{$M_{\ast} / M_{\odot}$\tablenotemark{a}} &
\colhead{$R_{\ast} / R_{\odot}$\tablenotemark{a}} &
\colhead{$L_{\ast} / L_{\odot}$\tablenotemark{a}} &
\colhead{(G)} &
\colhead{($M_{\odot} / \mbox{yr}$)} &
\colhead{($M_{\odot} / \mbox{yr}$)} &
\colhead{$\delta$}
}

\startdata

 AA Tau &
5.81& 0.38& 1.8 &  0.6457& 2780 & --6.9 & --9.1  & 0.006 \\
 &
5.98& 0.53& 1.74&  0.7079& 2780 & --8.48& --9.1  & 0.006 \\
 BP Tau &
5.78& 0.45& 1.9 &  0.8710& 2170 & --6.8 & \ldots & 0.007 \\
 &
5.79& 0.49& 1.99&  0.9333& 2170 & --7.54& \ldots & 0.007 \\
 CY Tau &
6.27& 0.58& 1.4 &  0.4677& 1160 & --8.2 & --10.0 & 0.041 \\
 &
6.32& 0.42& 1.63&  0.4898& 1160 & --8.12& --10.0 & 0.041 \\
 DE Tau &
4.74& 0.24& 2.7 &  1.0471& 1120 & --6.5 & --9.2  & 0.040 \\
 &
5.02& 0.25& 2.45&  0.8710& 1120 & --7.58& --9.2  & 0.040 \\
 DF Tau &
3.74& 0.17& 3.9 &  2.0417& 2900 & --5.9 & --8.3  & 0.023 \\
 &
5.09& 0.27& 3.37&  1.3490& 2900 & --6.91& --8.3  & 0.023 \\
 DK Tau &
5.30& 0.38& 2.7 &  1.6982& 2640 & --6.4 & --8.5  & 0.002 \\
 &
5.56& 0.43& 2.49&  1.4454& 2640 & --7.42& --8.5  & 0.002 \\
 DN Tau &
5.65& 0.42& 2.2 &  1.0715& 2000 & --7.5 & --9.4  & 0.005 \\
 &
5.69& 0.38& 2.09&  0.8710& 2000 & --8.46& --9.4  & 0.005 \\
 DO Tau &
5.21& 0.31& 2.4 &  1.1220& 1989\tablenotemark{b} & --5.6 & --7.5  & 0.088 \\
 &
5.63& 0.37& 2.25&  1.0000& 2164\tablenotemark{b} & --6.84& --7.5  & 0.088 \\
 DS Tau &
6.54& 1.28& 1.6 &  1.3490& 2714\tablenotemark{b} & --6.6 & \ldots & 0.005 \\
 &
6.58& 0.87& 1.36&  0.5754& 3119\tablenotemark{b} & --7.89& \ldots & 0.005 \\
 GG Tau &
4.82& 0.29& 2.8 &  1.4791& 1240 & --6.7 & --9.1  & 0.014 \\
 &
5.63& 0.44& 2.31&  1.2589& 1240 & --7.76& --9.1  & 0.014 \\
 GI Tau &
5.11& 0.30& 2.5 &  1.1749& 2730 & --6.9 & --9.4  & 0.004 \\
 &
6.09& 0.71& 1.48&  0.8511& 2730 & --8.02& --9.4  & 0.004 \\
 GM Aur &
5.06& 0.52& 1.6 &  1.6982& 2220 & --7.6 & \ldots & 0.001 \\
 &
5.95& 0.52& 1.78&  0.7413& 2220 & --8.02& \ldots & 0.001 \\
 HN Tau &
6.86& 0.72& 1.1 &  0.2754& 3860\tablenotemark{b} & --7.7 &
 --8.34\tablenotemark{c} & 0.009 \\
 &
7.48& 0.81& 0.76&  0.1905& 4198\tablenotemark{b} & --8.89&
 --8.34\tablenotemark{c} & 0.009 \\
 UY Aur &
4.93& 0.29& 2.6 &  1.3183& 1858\tablenotemark{b} & --6.6 & --8.2  & 0.010 \\
 &
5.52& 0.42& 2.60&  1.0471& 2273\tablenotemark{b} & --7.18& --8.2  & 0.010 \\

\enddata
\tablenotetext{a}{For each star, the values in the first row come
from Hartigan et al.\  (1995); values in the second row come from
Hartmann et al.\  (1998).  All logarithms are base-10.}
\tablenotetext{b}{Mean magnetic fields for these stars were estimated
by scaling with the equipartition field strength in photosphere
(see the text).}
\tablenotetext{c}{The mass loss rate from Hartigan et al.\  (1995)
was replaced with a slightly lower revised value from
Hartigan et al.\  (2004).}

\end{deluxetable*}

Table 1 lists 14 stars in Taurus-Auriga that have published
measurements of both the accretion rate $\dot{M}_{\rm acc}$ and
the accretion-spot filling factor $\delta$.
Since there were two independent determinations of the
fundamental stellar parameters and the accretion rates, each
star is listed with two rows that correspond to the two separate
sets of observations.
Differences between the two rows convey a sense of the
measurement uncertainties.
The first row uses tabulated ages $t$, masses $M_{\ast}$,
radii $R_{\ast}$, and luminosities $L_{\ast}$ from
Hartigan et al.\  (1995), which is denoted HEG.
The second row uses tabulated results from Hartmann et al.\  (1998),
which is denoted HCGD.
Measurements of other quantities, taken from elsewhere in the
literature, are duplicated in both rows for each star.
In a way, the full list of rows in Table 1 corresponds to 28
quasi-independent empirical stellar data points that can be
compared with the theoretical models.

Note that the stellar ages determined by HEG seem to be
systematically younger, and the accretion rates larger, than
those computed by HCGD.
These two effects are likely to be related to one another, since
they are both derived (in part) from the stellar luminosity.
HEG generally reports higher luminosities, which puts the stars
``higher up'' in the Hertzsprung-Russell diagram (i.e.,
implies they are younger) and also gives larger absolute values
for the accretion luminosity.
Gullbring et al.\  (1998) discussed these discrepancies in
terms of differences in how the interstellar reddening is
estimated, how the radiative transfer calculations are done,
and how the magnetospheric geometry is taken into account.

\begin{figure}
\epsscale{1.17}
\plotone{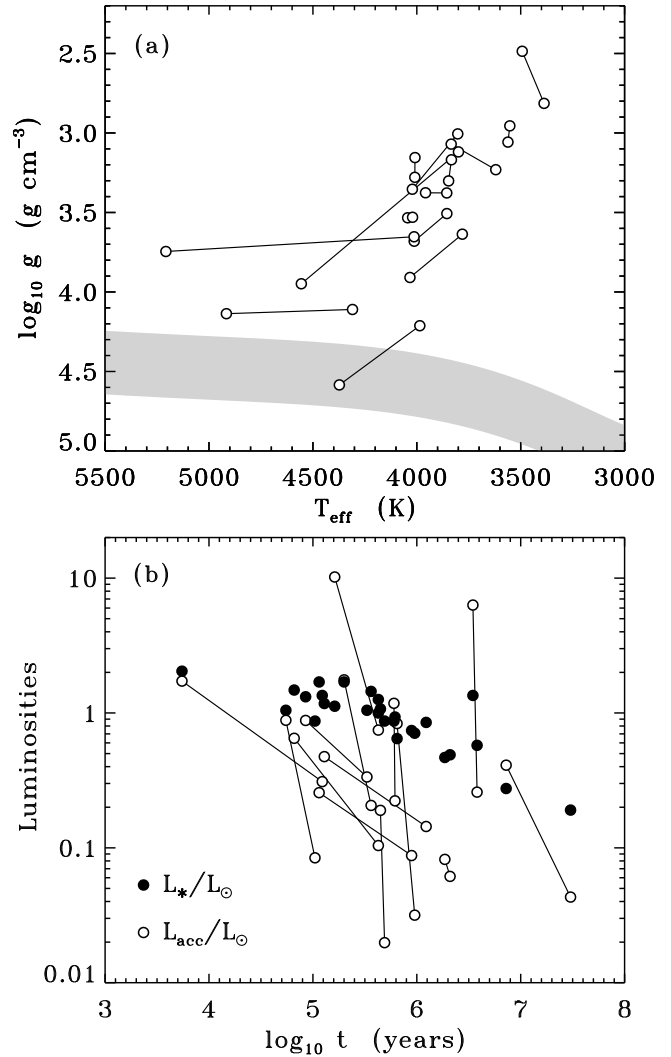}
\caption{Selection of derived stellar properties for the stars
listed in Table 1.
(a) Base-10 logarithm of surface gravity plotted against
effective temperature (open circles), compared with the
approximate position of the zero-age main sequence (gray band).
(b) Stellar photon luminosity (filled circles) and
accretion luminosity (open circles), both in units of the
solar luminosity, shown as a function of age.}
\end{figure}

Figure 1 illustrates some stellar properties derived from the
information given in Table 1.
As in many of the plots in this paper, straight lines join
together the independent determinations of quantities for each
star that were computed from the HEG and HCGD pairs of rows
in Table 1.
The effective temperature $T_{\rm eff}$ and surface gravity
$g = G M_{\ast}/R_{\ast}^{2}$ were computed assuming spherical
symmetry.
The accretion luminosity $L_{\rm acc}$ was estimated as
$G M_{\ast} \dot{M}_{\rm acc}/R_{\ast}$, which ignores an
order-unity correction factor for the free-fall from the
inner edge of the disk (see Gullbring et al.\  1998).
Several points in these plots that show anomalous values
deserve some extra mention.
\begin{enumerate}
\item
The hottest star in the sample, with $T_{\rm eff} = 5207$ K,
corresponds to the HEG measurement for GM Aur.
The HEG luminosity is about 2.3 times larger than the HCGD
luminosity, which could account for both the larger effective
temperature and its younger age.
Other sources (e.g., Gullbring et al.\  1998) report a luminosity
that agrees better with the HCGD value, and thus it may be wise to
ignore the HEG properties for this star.
\item
The coolest and lowest-gravity star in Figure 1(a) is DF Tau.
The HEG measurement has a slightly low gravity ($\log g = 2.49$)
when compared to other stars of similar age, but the HCGD
measurements also indicate that its gravity is lower than
average ($\log g = 2.81$).
The two sources agree about DF Tau being young (in age) and
large (in radius), but the HEG value of the age ($\log t = 3.74$)
appeared to involve {\em extrapolation} beyond the range of
ages used in the pre-main-sequence tracks of
D'Antona \& Mazzitelli (1994).
Although this may have yielded an incorrect age, it does not
affect the analysis of this paper, since the ages are not used
as input parameters for any of the models.
\item
There are two data points in Figure 1(b) with
$L_{\rm acc} \gg L_{\ast}$.
The HEG measurement of DO Tau (at $\log t = 5.21$) exhibits the
largest accretion rate of the sample in both sets of measurements,
as well as the largest mass outflow rate and the largest value
of $\delta$.
Thus, DO Tau appears to be a genuinely active object.
The other star with a large accretion luminosity is the HEG
measurement of DS Tau (at $\log t = 6.54$).
Its accretion rate is typical, but it has an anomalously high
ratio of $M_{\ast} / R_{\ast}$.
HCGD also list DS Tau as the most massive star of this sample,
but by less of a wide margin.
\end{enumerate}
It should also be noted that the spread of measured ages for
both the HEG and HCGD samples may be closely related to the
episodic nature of the accretion.
Intermittently variable accretion has been shown recently by
Baraffe et al.\  (2009) to produce a realistic spread of
luminosities, such that a population of stars with identical
ages $t \approx 1$ Myr may be interpreted (via standard
evolutionary tracks) to have an age spread of as much as 10 Myr.
In any case, the ages reported in Table 1 are used in this paper
only to organize the data in graphical form, and not as inputs
to the models.

The accretion rates given in Table 1 show a large degree of
intrinsic scatter, even for stars of similar age and mass (see
also Muzerolle et al.\  2003; Calvet et al.\  2004;
Nguyen et al.\  2009).
In the earliest stages of star formation, this scatter may be
the result of an inhomogeneous cluster environment
(Pfalzner et al.\  2008).
For CTTS, however, this observational scatter is also likely
to be the result of time variability in the three-dimensional
pattern of magnetospheric accretion streams that connect the
star and its surrounding disk.
In any case, the observed values of $\dot{M}_{\rm acc}$
represent snapshots of the accretion that should ideally be
paired with {\em simultaneous} measurements of the wind's
mass loss rate and X-ray activity.
The observations used in this paper are not tightly constrained
to be exactly contemporaneous, but in many cases the substantial
uncertainty in these values lessens the importance of this issue.

The fraction of stellar surface area $\delta$ covered by accretion
streams was computed for each star by Calvet \& Gullbring (1998).
They compared observed and modeled spectral energy distributions
(i.e., Balmer and Paschen continua) to match both the total
accretion energy flux and the emitting area on the stellar surface.
There was no clear trend with age in these determinations of
$\delta$, nor was there in the earlier measurements of
Valenti et al.\  (1993).
It is important to note that both other observational diagnostics
(Costa et al.\  2000; Batalha et al.\  2002; Donati et al.\  2007)
as well as some simulations (Romanova et al.\  2004)
have given rise to larger accretion-spot filling factors up to
10\% or more.
It is possible that these diagnostics may probe not just the
magnetic flux tubes filled by the densest clumps, but also
larger surrounding volumes that are energized by nearby
accretion impacts (see also Brickhouse et al.\  2009).

The surface magnetic field strengths $B_{\ast}$ shown in Table 1
were measured by Johns-Krull (2007) via the Zeeman broadening of
several \ion{Ti}{1} lines in the infrared.
Only 10 of the 14 stars were observed, and for the other four we
estimated their photospheric field strengths by assuming a mean
level of partitioning between gas pressure and magnetic pressure
(see below).

The mass loss rates $\dot{M}_{\rm wind}$ given in Table 1 were
computed by HEG from the properties of high-velocity blueshifted
emission components in forbidden lines such as [\ion{O}{1}]
$\lambda$6300 (see also Kwan \& Tademaru 1995;
Hartigan et al.\  2004).
These values have uncertainties of up to an order of magnitude.
HEG reported only upper limits for BP Tau, DS Tau, and GM Aur,
and these are not included in Table 1.
It is uncertain whether these measured ``jet'' outflows originate
on the stellar surface or in the accretion disk (see, e.g.,
Paatz \& Camenzind 1996; Calvet 1997; Ferreira et al.\  2006;
Azevedo et al.\  2007; Fendt 2009).
If the empirical mass loss rates represent measurements of the
outflow far from the stellar surface, then it is possible that
they probe the {\em sum} of the mass fluxes from winds rooted
both on the star and in the disk.
In that case, these rates can be assumed to be tentative
upper bounds for the stellar wind component.

The 28 sets of fundamental parameters from Table 1 were used to
compute the photospheric mass densities $\rho_{\ast}$ at the
surfaces of the T Tauri stars.
The relevant criterion is that the Rosseland mean optical depth
should have a value of approximately one (i.e., $\tau_{\rm R}
\approx \kappa_{\rm R} \rho_{\ast} H_{\ast} = 1$) at the
stellar photosphere.
The scale height $H_{\ast}$ is defined for simplicity as
\begin{equation}
  H_{\ast} \, = \, \frac{c_{s}^{2}}{\gamma g} \, = \,
  \frac{k_{\rm B} T_{\rm eff}}{g m_{\rm H}}
  \label{eq:Hast}
\end{equation}
where $c_s$ is the sound speed, $g$ is the surface gravity,
$\gamma = 5/3$ for an adiabatic gas, $k_{\rm B}$ is Boltzmann's
constant, and $m_{\rm H}$ is the mass of a hydrogen atom.
The Rosseland mean opacity $\kappa_{\rm R}$ was interpolated
as a function of temperature and density from the tables of
Ferguson et al.\  (2005).
Figure 2 shows these derived photospheric densities along with
other densities measured higher in the accretion streams that
are defined below.

\begin{figure}
\epsscale{1.17}
\plotone{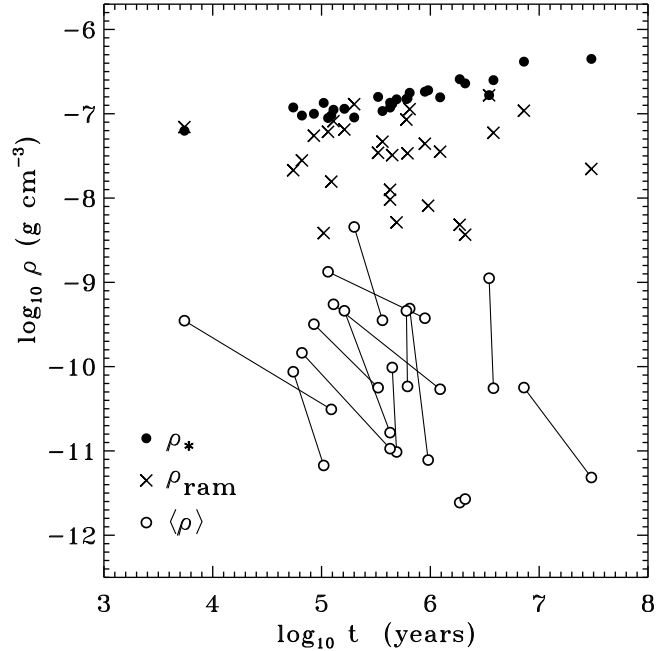}
\caption{Mass densities, plotted as a function of age, at the
stellar photosphere (filled circles), at the accretion shock
(crosses), and averaged over clumps and inter-clump regions in
the accretion streams (open circles).}
\end{figure}

For each star with a measured magnetic field strength, the
magnetic pressure $B_{\ast}^{2} / 8 \pi$ was compared to the
photospheric gas pressure $P_{\ast} = \rho_{\ast} k_{\rm B}
T_{\rm eff} / m_{\rm H}$.
The equipartition field strength $B_{\rm eq}$ is defined as
the field strength required to produce a magnetic pressure
equal to the gas pressure.
For the 10 stars (i.e., 20 rows of Table 1) for which this
comparison could be made, the resulting values of the ratio
$B_{\ast} / B_{\rm eq}$ ranged between 0.79 and 4.26.
The mean value of the ratio was 2.084 and its standard
deviation was 0.90.
Thus, for the remaining four stars without magnetic field
measurements, the magnetic field strength was estimated using
this mean value $B_{\ast} = 2.084 B_{\rm eq}$.

\vspace*{0.05in}
\section{Properties of Accretion Streams and Waves}

During the CTTS phase, accretion of matter is believed to
occur via supersonic ballistic infall along magnetic
flux tubes that thread the inner disk (see
Lynden-Bell \& Pringle 1974; Uchida \& Shibata 1984;
Camenzind 1990; K\"{o}nigl 1991).
The models developed in this paper make use of the idea that the
accretion streams are likely to be highly unstable and
time-variable, and thus much of the gas deposited onto the star
is expected to be in the form of intermittent clumps (e.g.,
Gullbring et al.\  1996; Safier 1998;
Bouvier et al.\  2003, 2004, 2007; Stempels 2003;
Giardino et al.\  2007).
Although most of the observed CTTS photometric variability
appears to be related to rotation of accretion streams and spots
(with timescales of days), there is also persistent low-level
hour-timescale variability that may reflect the internal
clumpiness of the streams.
Cranmer (2008) predicted that the number of accretion-related
flux tubes on the stellar surface is likely to be of order
$10^2$, such that if Poisson-like statistics dominated the
distribution of clumps, the relative variability would be
only of order 10\%.
This is within the realm of the observed hour-timescale
variability (see, e.g., Rucinski et al.\  2008;
Nguyen et al.\  2009).

As described by Scheurwater \& Kuijpers (1988) and Cranmer (2008),
the impact of each clump is expected to generate MHD waves and
turbulence on the stellar surface.
When these fluctuations propagate across the surface, they spread
out the kinetic energy of accretion and inject some of it into
the star's global magnetic field.
This analysis neglects possible additional motions that could
be driven by instabilities in the accretion shock itself (e.g.,
Chevalier \& Imamura 1982; Mignone 2005).
For strong enough magnetic fields on CTTS surfaces, these
instabilities appear to be suppressed (Koldoba et al.\  2008).
However, in some simulations there are long-term transitions
between periods of stability and instability
(Romanova et al.\  2008).

The dynamical properties of accretion streams are modeled here
assuming ballistic infall---from the inner edge of the accretion
disk---along field lines described by an axially symmetric
dipole (e.g., Calvet \& Hartmann 1992; Muzerolle et al.\  2001).
Although we know that the actual magnetic fields of T Tauri
stars can have non-dipolar components (Donati et al.\  2007;
Gregory et al.\  2008), the dipole assumption allows representative
geometrical properties of the streams to be calculated easily.
In the absence of explicit specification below, the parameters
and methods used to model the accretion streams are the same as
those of Cranmer (2008).

The field lines that thread the disk intercept the equatorial
plane between inner and outer radii $r_{\rm in}$ and $r_{\rm out}$.
The location of the inner radius (also called the truncation radius)
is specified here using K\"{o}nigl's (1991) assumption of pressure
balance between the accretion stream and the stellar magnetic field,
with
\begin{equation}
  r_{\rm in} \, = \, 0.5 \left( \frac{B_{\ast}^{4} R_{\ast}^{12}}
  {2G M_{\ast} \dot{M}_{\rm acc}^{2}} \right)^{1/7}  \,\, .
  \label{eq:rin}
\end{equation}
The outer radius can be computed if we know the fraction of the
stellar surface $\delta$ that is covered by the accretion streams.
For a dipole field, these quantities satisfy
\begin{equation}
  \sqrt{1 - \frac{R_{\ast}}{r_{\rm out}}} \, = \,
  \delta + \sqrt{1 - \frac{R_{\ast}}{r_{\rm in}}} \,\, .
  \label{eq:rout}
\end{equation}
A field line that intersects the disk at a radius $r > r_{\rm in}$
encounters the stellar surface at a colatitude $\theta$, where
$\sin^{2}\theta = R_{\ast}/r$.
Thus, the radii $r_{\rm in}$ and $r_{\rm out}$ correspond to surface
colatitudes $\theta_{\rm in}$ and $\theta_{\rm out}$, respectively.

For the stellar properties listed in Table 1, the ratio
$r_{\rm in} / R_{\ast}$ ranges between 1.3 and 8.
These values correspond roughly to the highest and lowest accretion
rates, respectively.
The ratio $r_{\rm out} / r_{\rm in}$ ranges between 1.006 and 1.81
(with a median value of 1.066).
The mean colatitudes of the accretion streams range from about
{20\arcdeg}, for the oldest stars with the lowest accretion rates,
to {60\arcdeg}, for the youngest stars with highest accretion rates.
The fractional surface area covered by the ``polar caps''
$\delta_{\rm pol}$ (i.e., the regions between colatitudes 0 and
$\theta_{\rm out}$ on each pole) does not vary significantly with
age, and its mean value for all of the stars is about 0.15.
The fractional surface area covered by the equatorial band of
closed magnetic fields (i.e., $\delta_{\rm equ} =
1 - \delta - \delta_{\rm pol}$) has an average value of about 0.83.
Although these numbers depend on the assumption of dipole symmetry,
they are more empirically grounded than the models of Cranmer (2008).
In that paper, both the ratio $r_{\rm out} / r_{\rm in}$ and the
surface field strength $B_{\ast}$ were specified as arbitrary
constants, but here they are based on specific measurements for
each star.

The accreting gas is accelerated from rest at the inner edge of
the disk to a ballistic free-fall speed at the stellar surface,
\begin{equation}
  v_{\rm ff} \, = \, \left[ \frac{2 G M_{\ast}}{R_{\ast}}
  \left( 1 - \frac{R_{\ast}}{r_{\rm in}} \right) \right]^{1/2}
  \,\, .
  \label{eq:vff}
\end{equation}
This expression slightly underestimates the mean free-fall velocity,
since the streams actually come from all radii between $r_{\rm in}$
and $r_{\rm out}$.
The free-fall speeds at the stellar surface range between 100 and
600 km s$^{-1}$, and they increase with increasing age roughly as
$t^{0.22}$ because of a similar behavior in the stellar escape speed.

In the models of Cranmer (2008) it was assumed that the free-fall
speed $v_{\rm ff}$ is identical to the speed of infalling clumps
$v_{c}$ that occupy the accretion streams.
Here, however, the wider range of stellar properties and accretion
rates (in Table 1) necessitates a closer look at that assumption.
Scheurwater \& Kuijpers (1988) found that the optimal situation
was for the clump's ram pressure to be larger than the gas pressure
in the stream, but smaller than the magnetic pressure.
When these conditions are not met, the clumps may be decelerated
(with $v_{c} < v_{\rm ff}$).
Considering only the ``stopping power'' of a large ambient gas
pressure, equation (64) of Scheurwater \& Kuijpers (1988) gave
an upper-limit clump speed of
\begin{equation}
  v_{c} \, \leq \, \left( \frac{2}{3} \frac{r_c}{R_{\ast}}
  \frac{\rho_c}{\rho_0} \right)^{1/2} v_{\rm ff}
  \label{eq:vc}
\end{equation}
where $r_{c} \approx R_{\ast} (\theta_{\rm in}-\theta_{\rm out})/2$
is the radius of a representative spherical clump at the surface,
$\rho_c$ is the density interior to each clump, and $\rho_0$
is the ambient density between clumps in the accretion streams.
The Cranmer (2008) models did not depend explicitly on the
ratio $\rho_{c} / \rho_{0}$, but here it needs to be computed.

A first estimate of the density contrast of supersonically
accelerated clumps may involve the fact that such features could
easily form {\em shocks} in the accretion streams.
In that case, the maximum density contrast would be of order
$(\gamma + 1)/(\gamma - 1) \approx 4$, with $\gamma = 5/3$ for
an adiabatic gas.
However, Elmegreen (1990) showed that MHD waves can form in this
kind of infall-dominated region and lead to the occurrence of
colliding nonlinear wave packets.
These can efficiently clear out the gas from the inter-clump
regions and thus enhance the density contrast up to factors of
order 100.
Note that if we assumed that $\rho_{c}/ \rho_{0} \rightarrow \infty$,
the upper limit in equation (\ref{eq:vc}) would never be applied.
On the other hand, assuming too low a value for the density
contrast ratio would lead to an unrealistic over-application of
this limit.
Thus, a conservative approach would be to use a value for this
ratio that may err on the side of being too large.
In the absence of other information, the trial value of 100 was
chosen.
This ends up reducing $v_c$ below $v_{\rm ff}$ in the majority
of cases (20 out of 28) by about a factor of two, and allowing
$v_{c} = v_{\rm ff}$ in the other 8 cases.

The other possible deceleration effect involves a comparison
between the clump's ram pressure and the magnetic pressure of
the accretion stream.
The theoretical development of Scheurwater \& Kuijpers (1988)
and Cranmer (2008) assumed that the passage of a clump does not
disrupt the mean magnetic field in the stream.
However, if a clump's ram pressure begins to exceed the local
magnetic pressure, MHD instabilities may occur that further
decelerate the clump to a point where it can no longer disrupt
the field.
Thus, for each star, we made sure that $v_c$ never exceeds
half of the intra-clump Alfv\'{e}n speed
$B_{0} / (4\pi\rho_{c})^{1/2}$ (see equation [1] of
Scheurwater \& Kuijpers 1988).
This further reduction was applied in 10 out of the 28 cases,
and the final saturated values of $v_c$ were found to range
between about 30 and 530 km s$^{-1}$.

The accretion streams are assumed to be stopped at the location
where the ram pressure,
\begin{equation}
  P_{\rm ram} \, = \, \frac{\rho v_{c}^2}{2} \, = \,
  \frac{v_{c} \dot{M}_{\rm acc}}{8 \pi \delta R_{\ast}^2}
  \,\, ,
  \label{eq:Pramdef}
\end{equation}
is balanced by the gas pressure of the undisturbed stellar atmosphere
(e.g., Hartmann et al.\  1997; Calvet \& Gullbring 1998).
This paper uses a new algorithm to estimate the density
$\rho_{\rm ram}$ at which this balance occurs.
The approximation given by Cranmer (2008)---who used the notation
$\rho_{\rm sh}$ for this quantity---assumed that the ram pressure
balance occurs well above the photosphere, such that the temperature
has reached its minimum value of approximately $\sim 0.84 T_{\rm eff}$
(for radiative equilibrium in a gray atmosphere).
However, larger accretion rates imply that the accretion shock may
penetrate deeper into higher-temperature regions.
The new prescription uses a gray atmosphere relation for
temperature versus optical depth (Mihalas 1978), with
\begin{equation}
  T^{4} \, = \, \frac{3}{4} T_{\rm eff}^{4} \left( \tau +
  \frac{2}{3} \right)
\end{equation}
and the height dependence of the optical depth $\tau$ is approximated
via hydrostatic equilibrium and a constant opacity; i.e.,
$\tau(z) \approx \kappa \rho(z) H_{\ast}$.
In this case, the ram pressure balance condition is given by
\begin{equation}
  P_{\rm ram} \, = \, 0.84 P_{\ast}
  e^{-z_{\rm ram}/H_{\ast}} \left( 1 + e^{-z_{\rm ram}/H_{\ast}}
  \right)^{1/4}
  \label{eq:Prambal}
\end{equation}
where $z_{\rm ram}$ is the height of ram pressure balance that
must be solved for numerically.
Once $z_{\rm ram}$ has been determined, the density at this height
is given straightforwardly by
$\rho_{\rm ram} = \rho_{\ast} e^{-z_{\rm ram}/H_{\ast}}$.
The approximate version given in equation (10) of Cranmer (2008)
is equivalent to ignoring the final exponential term in
equation (\ref{eq:Prambal}) above.
Figure 2 shows $\rho_{\rm ram}$ for the stars modeled in this paper.
Note that the shock penetrates below the photosphere
($\rho_{\rm ram} > \rho_{\ast}$) for only two of the 28 data
points; these are the HEG values of DF Tau and DK Tau.

The energy in Alfv\'{e}n waves released on the stellar surface from
a single clump impact was given by Scheurwater \& Kuijpers (1988) as
\begin{equation}
  E_{\rm A} \, = \, 0.06615 \,\, \frac{\rho_c}{\rho_0}
  \left( \frac{v_c}{V_{\rm A}} \right)^{3} m_{c} v_{c}^{2}
  \label{eq:EA}
\end{equation}
where $m_c$ is the mass in the clump and the Alfv\'{e}n speed
$V_{\rm A}$ is defined as that of the ambient medium, with
$V_{\rm A}  = B_{0}/(4\pi\rho_0)^{1/2}$.
Because these equations are applied near the stellar surface,
the magnetic field strength in the accretion stream $B_{0}$ is
assumed to be equal to $B_{\ast}$.
Cranmer (2008) worked out the conversion from a single clump's
wave energy yield to the time-steady amplitude of Alfv\'{e}n
waves (from the continuous accretion of many clumps) at the
north and south poles of a star with a dipole field geometry.
Using equations (19)--(22) of Cranmer (2008), the transverse
velocity amplitude of waves in the polar photosphere can be
written as
\begin{equation}
  v_{\perp \ast} \, = \, 0.5145 \left(
  \frac{\langle \rho \rangle \rho_{\rm ram} \delta}
  {\rho_{c} \rho_{\ast} \Delta\theta} \right)^{1/2}
  \frac{v_{c}^3}{B_{\ast}^{2} / (4\pi \rho_{c})}
  \label{eq:vperpast}
\end{equation}
where $\Delta \theta = (\theta_{\rm in} - \theta_{\rm out})
(\theta_{\rm in} + \theta_{\rm out})$ and the time-averaged
mean density in the accretion streams is
\begin{equation}
  \langle \rho \rangle \, = \, \frac{\dot{M}_{\rm acc}}
  {4 \pi \delta R_{\ast}^{2} v_{c}}  \,\, .
  \label{eq:rhoavdef}
\end{equation}
This quantity is intermediate between $\rho_0$ and $\rho_c$, and
it is plotted in Figure 2 for the 28 stellar data points.
The total flux of Alfv\'{e}n waves at the pole is given by
$F_{\rm A} = \rho_{\ast} v_{\perp \ast}^{2} V_{{\rm A} \ast}$,
where $V_{{\rm A} \ast}  = B_{\ast}/(4\pi\rho_{\ast})^{1/2}$.
The fluxes of waves that reach other regions on the stellar
surface (away from the poles) are discussed in Sections 4 and 6.

\begin{figure}
\epsscale{1.17}
\plotone{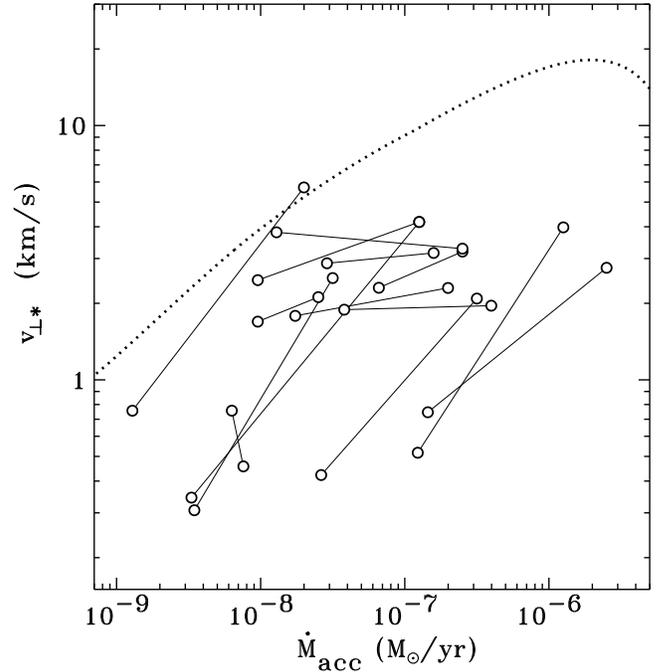}
\caption{Photospheric velocity amplitudes of Alfv\'{e}n waves induced
by magnetospheric accretion, plotted as a function of the accretion
rate.  Wave velocities were computed at the north and south poles for
the stars listed in Table 1 (open circles) and for the model
evolutionary sequence of a 1 $M_{\odot}$ star from Cranmer (2008)
(dotted line).}
\end{figure}

Figure 3 shows the Alfv\'{e}n wave velocity amplitude for each
star as a function of the accretion rate.
Also shown is the continuous relationship for the idealized
evolutionary sequence of a 1 $M_{\odot}$ star from Cranmer (2008).
The wave amplitudes derived for the current stellar sample are
generally lower than those from Cranmer (2008).
Part of this is due to the new models taking into account the two
ram-pressure-deceleration effects that were discussed above.
Also, it should be noted that the average stellar mass for the
stars in Table 1 is less than 1 $M_{\odot}$, so even the
unmodified free-fall speeds $v_{\rm ff}$ were generally lower
than those of Cranmer (2008).

\section{Stellar Winds from Polar Regions}

For several decades, MHD waves have been studied as likely sources
of energy and momentum for accelerating winds from cool stars
(see, e.g., Hollweg 1978; Hartmann \& MacGregor 1980; DeCampli 1981;
Wang \& Sheeley 1991; Airapetian et al.\  2000;
Falceta-Gon\c{c}alves et al.\  2006; Suzuki 2007).
The models of polar T Tauri winds developed by Cranmer (2008)
involved two sources for the MHD waves at the photosphere:
(1) convection-generated waves that propagate up from the interior,
and (2) accretion-generated waves of the form discussed in Section 3.
The Alfv\'{e}n waves were assumed to propagate up the open field
lines, dissipate via an MHD turbulent cascade, and heat the plasma
over an extended range of distance.
The resulting winds that had low (solar-type) mass loss rates were
driven by a combination of coronal gas pressure and wave pressure.
The high mass-loss winds of the youngest stars, however, were
radiatively cooled to chromospheric ($T \sim 10^{4}$ K) temperatures
and thus were dominated by wave pressure.
This paper makes the assumption that the T Tauri stars of Table 1
correspond to the latter physical regime of cool, wave-driven winds.

It was not possible to simply apply the cool-wind modeling technique
of Cranmer (2008) in this paper.
That methodology depended on knowing the exact values for several
parameters that cannot be reliably estimated for the present database
of T Tauri stars.
For example, many of the details of the numerical wind models
depended on the photospheric filling factor of magnetic flux
tubes.
The models of Cranmer (2008) used a low filling factor that is
appropriate for the present-day Sun, but possibly not for young CTTS.
Furthermore, the analytic young-star model described in
Sections 5.2 and 5.3 of Cranmer (2008) relied on an empirical
extrapolation of the Alfv\'{e}n wave amplitude at the critical
point from the numerical models of older stars in the
evolutionary sequence.
The general cool wave-driven wind theory of Holzer et al.\  (1983)
thus needed to be reanalyzed and adapted for the observational
stellar sample discussed above.

Before discussing the wind model, however, there is one modification
of the lower boundary condition on $v_{\perp \ast}$ that should
be specified.
Equation (\ref{eq:vperpast}) gives the photospheric Alfv\'{e}n wave
amplitude at the north or south pole of the magnetic stellar dipole.
Using this value would give the properties of the stellar wind
only over the poles.
But in order to compute the {\em total} mass loss rate
$\dot{M}_{\rm wind}$ that comes from the open fields spanning the
two polar cap regions, we decided to average over these polar caps.
The wave flux from equation (\ref{eq:vperpast}) is thus multiplied
by the following correction factor
\begin{equation}
  C_{\rm pol} \, = \, \frac{1}{\delta_{\rm pol}}
  \int_{0}^{\theta_{\rm out}} d\theta \, \sin\theta \,\,
  \frac{F_{\rm A} (\theta)}{F_{\rm A} (0)}  \,\, .
  \label{eq:Cpol}
\end{equation}
The expressions given in Section 3.4 of Cranmer (2008) are required
to compute the latitude dependent wave flux $F_{\rm A} (\theta)$.
The polar value of $v_{\perp \ast}$ is multiplied by
$C_{\rm pol}^{1/2}$ to produce the mean value for the entire
polar-cap region.

\begin{figure}
\epsscale{1.17}
\plotone{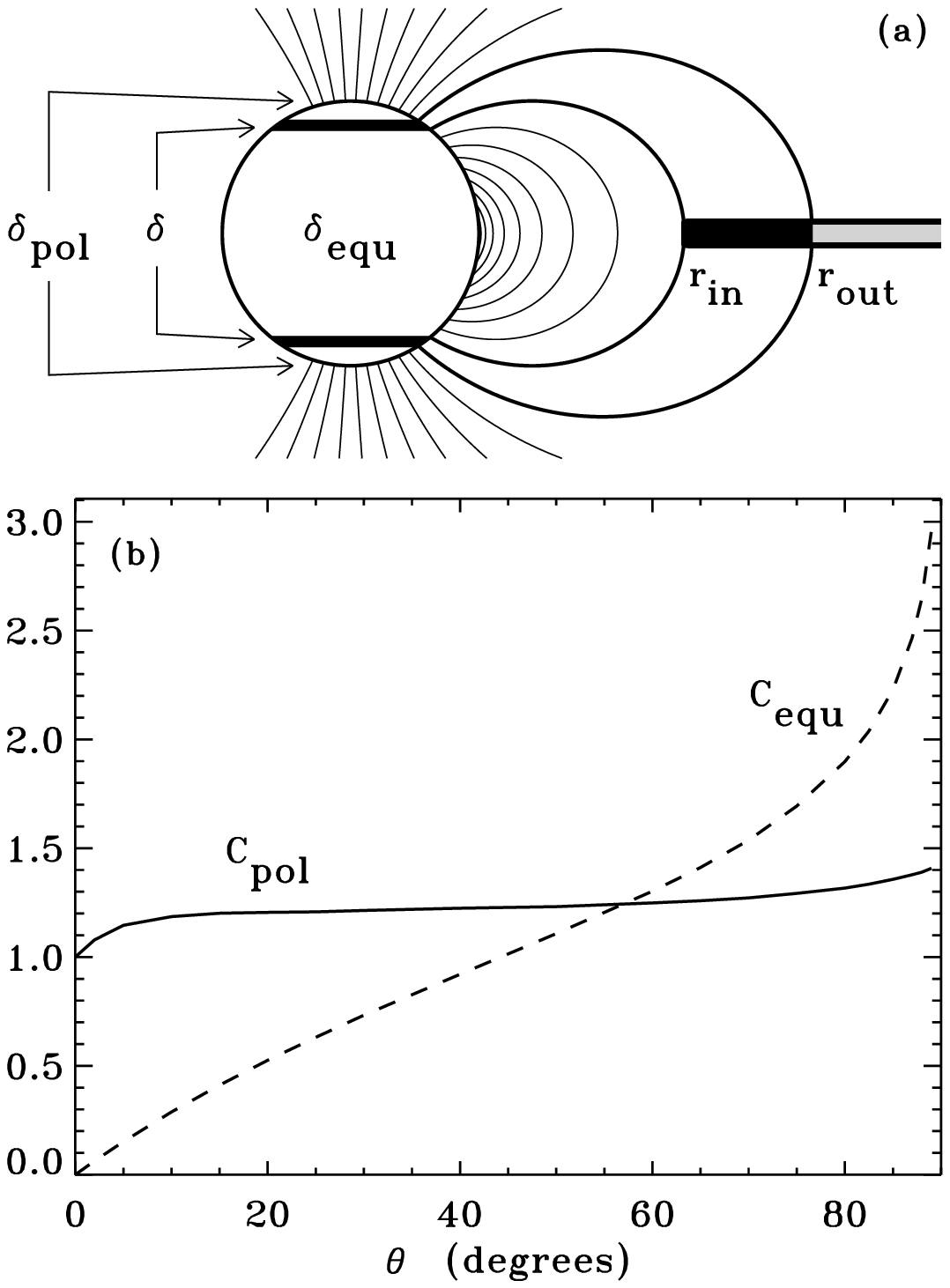}
\caption{(a) Modeled magnetic field geometry, showing
inner and outer equatorial radii $r_{\rm in}$ and $r_{\rm out}$
between which accretion takes place, as well as fractional
areas on the stellar surface subtended by accretion streams
($\delta$), open-field polar caps ($\delta_{\rm pol}$), and
equatorial closed-field loops ($\delta_{\rm equ}$).
(b) Correction factors for photospheric Alfv\'{e}n wave
fluxes plotted versus the mean colatitude $\theta$ of accretion
streams on the surface.  $C_{\rm pol}$ and
$C_{\rm equ}$ specify ratios of surface-averaged wave
flux (with the averages taken over the polar and equatorial
regions, respectively) to the wave flux at the poles.}
\end{figure}

Figure 4 illustrates the relevant geometry and how $C_{\rm pol}$
varies as a function of the mean polar-cap latitude
$(\theta_{\rm in} + \theta_{\rm out})/2$.
Note that the application of this correction factor gives rise to
slightly larger wave amplitudes than those shown in Figure 3.
The averaging over the entire polar cap includes points on the
stellar surface that are {\em closer} to the accretion ring
which have larger local wave amplitudes.
However, the magnitude of the $C_{\rm pol}$ correction factor
is never far from unity, and thus it gives rise to only a
relatively small change in $v_{\perp \ast}$.
It is included here for consistency with the use of a corresponding
equatorial correction factor $C_{\rm equ}$ that is more important
when computing the properties of the closed low-latitude loops
(see Section 6).

The calculation of the stellar wind's mass loss rate largely
follows the development of Holzer et al.\  (1983).
Two key assumptions are: (1) that the Alfv\'{e}n wave amplitudes
are larger than the sound speeds in the wind, and (2) that there
is negligible wave damping between the stellar surface and the
wave-modified ``critical point'' of the flow.
A third assumption from Holzer et al.\  (1983)---which was initially
not applied but later found to be valid---is that the stellar
wind is sub-Alfv\'{e}nic at the critical point (i.e., that the wind
speed is much smaller than $V_{\rm A}$ at the critical point).
A first round of numerical models was constructed without the
third assumption.
These models integrated the equation of motion up and down from
the critical point, and used an iteration technique to determine
the true critical point radius $r_{\rm crit}$.
However, the resulting values of $r_{\rm crit}$ always remained
close to the numbers given by equation (35) of
Holzer et al.\  (1983), in which all three of the above
assumptions were applied.
This expression,
\begin{equation}
  \frac{r_{\rm crit}}{R_{\ast}} \, \approx \, 1 + \frac{3}{2}
  \left( \frac{d \ln A}{d \ln r} \right)^{-1}  \,\, ,
  \label{eq:rcrit}
\end{equation}
was thus used in the models described below, where $A(r)$
describes how the cross-sectional area of the open magnetic
flux tubes varies with distance.
For a dipole geometry over the poles, $A \propto r^{3}$ and
$r_{\rm crit} \approx 1.5 \, R_{\ast}$.
The current models assume that the filling factor of magnetic
flux tubes at the photosphere is unity (see, e.g., Saar 2001).

Once the critical point radius is known, it becomes possible to use
the same kinds of arguments as used in Section 5.3 of Cranmer (2008)
to estimate the wind velocity and density at the critical point,
and thus to compute the mass loss rate.
There are three unknown quantities and three equations to
constrain them.
The three unknowns are the critical point values of the wind
speed $u$, density $\rho$, and wave amplitude $v_{\perp}$.
The first equation is the constraint that the right-hand side of
the time-steady momentum equation must sum to zero at the
critical point of the flow (see, e.g., Parker 1958, 1963).
For the conditions described above, this gives
\begin{equation}
  \frac{GM_{\ast}}{r_{\rm crit}^2} \, = \,
  \frac{3 u_{\rm crit}^{2}}{r_{\rm crit}}
\end{equation}
and it is solved straightforwardly for $u_{\rm crit}$.
The second and third equations are, respectively, the definition
of the critical point velocity (in the ``cool'' limit of zero
gas pressure)
\begin{equation}
  u_{\rm crit}^{2} \, = \, \frac{v_{\perp}^2}{4}
  \left( \frac{1 + 3 M_{\rm A}}{1 + M_{\rm A}} \right)
  \label{eq:ucdef}
\end{equation}
and the conservation of wave action
\begin{equation}
  \rho v_{\perp}^{2} V_{\rm A} (1 + M_{\rm A})^{2} A \, = \,
  \mbox{constant}
  \label{eq:action}
\end{equation}
where $M_{\rm A} = u / V_{\rm A}$ is the Alfv\'{e}n Mach number
(see, e.g., Jacques 1977).
The constant in equation (\ref{eq:action}) is known for each star
because the conditions at the photosphere are known (and it is also
valid to assume $M_{\rm A} \rightarrow 0$ there as well).
The wave amplitude $v_{\perp}$ in equations (\ref{eq:ucdef}) and
(\ref{eq:action}) is evaluated at the critical point, where the
conservation of wave action gives rise to much larger values than
those at the photosphere (i.e., $v_{\perp} \gg v_{\perp \ast}$).

The fact that $V_{\rm A}$ itself depends on the density makes it
difficult to find an explicit analytic solution for
$\rho_{\rm crit}$.
Instead, the model code used here starts with a reasonable initial
guess and solves equations (\ref{eq:rcrit})--(\ref{eq:action})
iteratively until consistency is reached.
The resulting iterated values for $v_{\perp}$ at the critical
point range between 85 and 425 km s$^{-1}$.
In all cases, these Alfv\'{e}n wave amplitudes are larger than
the sound speeds in a cool chromospheric wind.
The resulting values for $\rho_{\rm crit}$ span more than five
orders of magnitude between $10^{-18}$ and
$4 \times 10^{-13}$ g cm$^{-3}$.

The stellar wind's mass loss rate was determined by
applying the equation of mass conservation and taking into account
that the open fields over the polar caps have {\em widened} by
the time that the wind reaches $r_{\rm crit}$.
For a dipole geometry, a rough estimate of this widening factor
is $\delta_{\rm crit} \approx \delta_{\rm pol}
(r_{\rm crit} / R_{\ast})^3$.
However, for two out of the 28 cases from Table 1, this
approximation gives a number greater than 1.
For these two cases the value of $\delta_{\rm crit}$ was
simply assumed to be 1.
Thus, the mass loss rate is estimated as
$\dot{M}_{\rm wind} = 4\pi r_{\rm crit}^{2} \delta_{\rm crit}
u_{\rm crit} \rho_{\rm crit}$.

\begin{deluxetable*}{ccccccccc}
\tablecaption{Modeled Wind, Shock, \& Coronal Loop Properties}
\tablewidth{0pt}

\tablehead{
\colhead{} &
\colhead{log $\dot{M}_{\rm wind}$} &
\colhead{log $T_{\rm sh}$} &
\colhead{log $n_{\rm post}$} &
\colhead{$\langle L \rangle$} &
\colhead{log $T_{\rm max}$\tablenotemark{a}} &
\colhead{log $n_{\rm TR}$\tablenotemark{a}} &
\colhead{log $n_{\rm top}$\tablenotemark{a}} &
\colhead{}
\\
\colhead{Object} &
\colhead{($M_{\odot} / \mbox{yr}$)} &
\colhead{(K)} &
\colhead{(cm$^{-3}$)} &
\colhead{(Mm)} &
\colhead{(K)} &
\colhead{(cm$^{-3}$)} &
\colhead{(cm$^{-3}$)} &
\colhead{$\alpha_{\rm R}$\tablenotemark{a}}
}

\startdata

 AA Tau &
--8.61  & 5.163 & 15.04 &  28.9 &  6.99 &  13.60 &  10.61 &  1.74 \\
 &
--13.66 & 5.892 & 13.27 &  18.6 &  6.75 &  12.86 &  10.11 &  1.98 \\
 BP Tau &
--9.10  & 4.986 & 15.01 &  27.0 &  6.91 &  13.36 &  10.45 &  1.77 \\
 &
--9.50  & 5.662 & 14.14 &  25.6 &  6.90 &  13.32 &  10.42 &  1.80 \\
 CY Tau &
--12.24 & 6.170 & 12.76 &  12.3 &  6.57 &  12.32 &   9.75 &  1.95 \\
 &
--12.79 & 5.982 & 12.80 &  22.0 &  6.58 &  12.35 &   9.78 &  1.97 \\
 DE Tau &
--8.90  & 5.125 & 14.29 &  94.4 &  6.82 &  13.09 &  10.27 &  1.77 \\
 &
--12.19 & 5.583 & 13.20 &  74.3 &  6.70 &  12.72 &  10.02 &  1.96 \\
 DF Tau &
--7.50  & 5.194 & 14.90 & 235.2 &  7.16 &  14.11 &  10.95 &  1.68 \\
 &
--11.68 & 5.521 & 13.86 & 105.4 &  7.00 &  13.63 &  10.63 &  1.95 \\
 DK Tau &
--9.55  & 4.305 & 15.84 &  52.0 &  7.00 &  13.63 &  10.63 &  1.82 \\
 &
--10.04 & 5.053 & 14.90 &  41.5 &  6.97 &  13.54 &  10.57 &  1.85 \\
 DN Tau &
--9.55  & 5.416 & 14.36 &  34.5 &  6.90 &  13.32 &  10.42 &  1.81 \\
 &
--13.46 & 5.585 & 13.36 &  35.6 &  6.73 &  12.81 &  10.08 &  1.98 \\
 DO Tau &
--8.70  & 4.916 & 15.00 &  58.2 &  6.95 &  13.49 &  10.53 &  1.74 \\
 &
--11.62 & 5.745 & 13.59 &  40.5 &  6.86 &  13.21 &  10.34 &  1.93 \\
 DS Tau &
--9.79  & 4.810 & 15.37 &   7.9 &  6.88 &  13.25 &  10.38 &  1.80 \\
 &
--9.78  & 5.955 & 14.12 &   7.4 &  6.90 &  13.32 &  10.42 &  1.81 \\
 GG Tau &
--8.96  & 4.997 & 14.51 &  84.8 &  6.85 &  13.16 &  10.32 &  1.77 \\
 &
--10.13 & 5.835 & 13.40 &  41.9 &  6.77 &  12.92 &  10.16 &  1.84 \\
 GI Tau &
--8.28  & 5.100 & 15.09 &  62.1 &  7.05 &  13.77 &  10.72 &  1.72 \\
 &
--10.52 & 5.767 & 14.10 &  10.0 &  6.87 &  13.24 &  10.37 &  1.86 \\
 GM Aur &
--10.56 & 4.594 & 15.41 &  18.2 &  6.87 &  13.23 &  10.36 &  1.85 \\
 &
--10.64 & 4.992 & 14.92 &  19.6 &  6.84 &  13.16 &  10.31 &  1.89 \\
 HN Tau &
--9.01  & 6.176 & 14.13 &   6.4 &  6.96 &  13.51 &  10.55 &  1.76 \\
 &
--13.41 & 6.570 & 13.06 &   2.2 &  6.70 &  12.73 &  10.03 &  1.98 \\
 UY Aur &
--8.53  & 5.008 & 14.85 &  73.8 &  6.95 &  13.48 &  10.53 &  1.73 \\
 &
--9.37  & 5.637 & 14.12 &  46.6 &  6.95 &  13.46 &  10.52 &  1.81 \\

\enddata
\tablenotetext{a}{Coronal loop quantities are presented for the
most-probable loop length $\langle L \rangle$ and the models that
use the iterated value of $\alpha_{\rm R}$ from equation
(\ref{eq:alrapp}).}

\end{deluxetable*}

Table 2 gives the resulting mass loss rates for the 28 stars in
the observational database, and Figure 5(a) plots the ratio
$\dot{M}_{\rm wind} / \dot{M}_{\rm acc}$ as a function of age.
The observationally determined mass loss rates from blueshifted
[\ion{O}{1}] $\lambda$6300 lines (HEG) are also shown, as are
the model results for the idealized evolutionary sequence of a
1 $M_{\odot}$ star (Cranmer 2008).
Of the 22 cases (11 stars) where it is possible to compare
directly with measured mass loss rates, 7 of the computed
values are within an order of magnitude of the observed rates.
The majority of the cases (17 out of 22) have lower modeled
rates than the observations, and only 5 have higher values.
For all 22 cases, the {\em median} value of the ratio of modeled
to measured $\dot{M}_{\rm wind}$ is 0.08.
As discussed further below, this implies that the outflows probed
by the [\ion{O}{1}] diagnostic may be dominated by the disk wind
component, which is not modeled here.

\begin{figure}
\epsscale{1.17}
\plotone{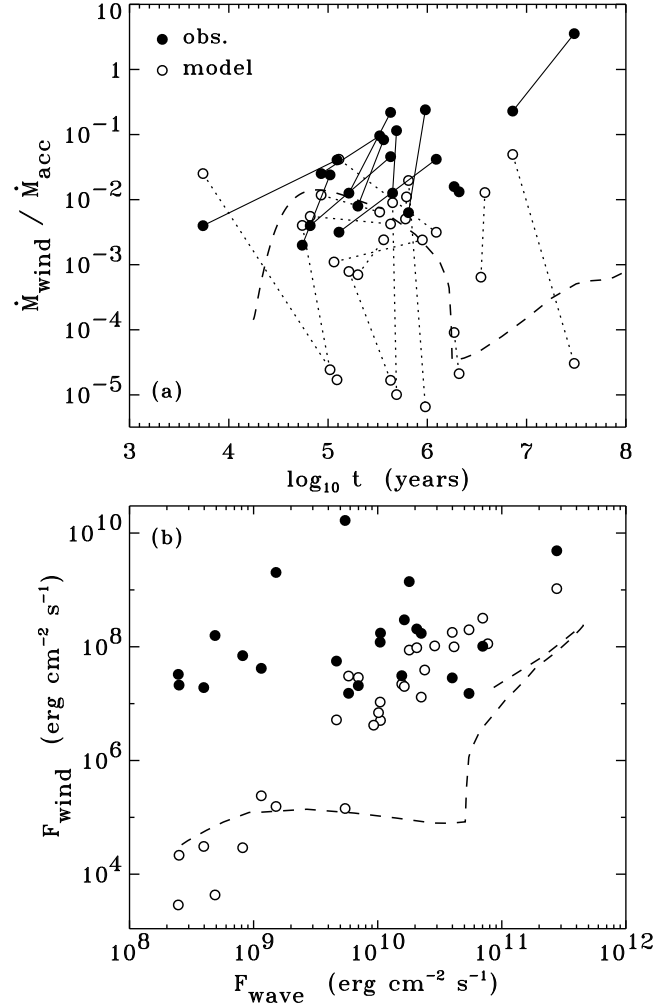}
\caption{(a) Stellar wind mass loss rates, normalized by the
accretion rates for each star, plotted versus age.
(b) Energy flux in the stellar wind plotted against the energy
flux in Alfv\'{e}n waves on the stellar surface.
The model calculations of this paper (open circles) are compared
with the HEG observations (filled circles) and the modeled
evolutionary sequence of Cranmer (2008) (dashed line).}
\end{figure}

Figure 5(b) compares the input flux of Alfv\'{e}n waves at
the photosphere with the computed energy flux of the stellar wind.
The Alfv\'{e}n wave flux $F_{\rm A} = F_{\rm wave}$ was defined
in Section 3, and the wind's energy flux is estimated to be
\begin{equation}
  F_{\rm wind} \, \approx \,
  \frac{L_{\rm wind}}{4\pi R_{\ast}^{2}}
  \, = \, \frac{GM_{\ast} \dot{M}_{\rm wind}}{4\pi R_{\ast}^{3}}
  \,\, .
  \label{eq:Fwind}
\end{equation}
The above quantity includes only the energy required to lift the
wind out of the star's gravity well, and not to accelerate it up
to its asymptotic terminal speed.
Thus it is not the {\em total} energy flux of the wind.
Note also that equation (\ref{eq:Fwind}) does not use any of
the polar-cap area factors defined above (e.g., $\delta_{\rm pol}$
or $\delta_{\rm crit}$) because the wind luminosity $L_{\rm wind}$
is presumably defined at an infinite distance where the flux
tubes have opened up to fill most of the circumstellar volume.
However, as long as a consistent definition is used for the
various quantities plotted in Figure 5(b), they can be
compared with one another fairly.

The fact that the modeled values of $F_{\rm wind}$ show a clear
correlation with $F_{\rm wave}$ in Figure 5(b) is not
surprising, since these winds are assumed to be wave-driven.
The dashed curve, from the evolutionary sequence of Cranmer (2008),
also shows a similar correlation.\footnote{%
The dip in the curve at intermediate wave fluxes between $10^{10}$
and $10^{11}$ erg cm$^{-2}$ s$^{-1}$ is most likely due to the
effects of heat conduction and radiative losses extracting a
fraction of the available energy that was liberated by MHD
turbulence; see Section 5.1 of Cranmer (2008).}
The wind fluxes from Cranmer (2008) are generally lower than
those computed in this paper because the former models used a
relatively small filling factor for the magnetic flux tubes at
the photosphere.
Smaller filling factors lead to more {\em dilution} of the wave
energy between the surface and the critical point of the wind,
and thus to less energy deposition overall.

In Figure 5(b), the model data points appear to be subdivided
into two groups, with a diving line around
$F_{\rm wind} \approx 10^{6}$ erg cm$^{-2}$ s$^{-1}$.
The group with the larger wind fluxes has a lower limit that is
close to the lower limit of the observed values.
Out of the 28 cases, 20 of the modeled wind fluxes fall into this
upper group, and only 8 are in the group that falls well short
of the observationally constrained wind fluxes.
At first glance, this may appear inconsistent with the fact
that approximately 77\% of the data points shown in Figure 5(a)
(i.e., 17 out of 22) have modeled ratios 
$\dot{M}_{\rm wind} / \dot{M}_{\rm acc}$ that are lower
than the observed ratios.
However, the 77\% fraction was computed by comparing each
model data point individually with its corresponding observation.
The corresponding 29\% fraction above (8 models out of 28 that
under-predict the observations) was found by examining the
overall distribution of points in Figure 5(b) and not matching
up the results for each star.
Also, the wind flux does not depend strongly on the accretion
rate, but $\dot{M}_{\rm acc}$ appears explicitly in the
denominator of the ratio shown in Figure 5(a).
The measured accretion rates show a large intrinsic range of
variability that influences the overall statistical agreement
or disagreement between the models and observations.

As also found by Cranmer (2008), many of the predicted mass loss
rates fall below those determined from the [\ion{O}{1}]
observations of HEG.
However, the level of limited agreement with the data is
interesting, because the measured mass loss rates are widely
believed to sample the much larger-scale disk wind or X-wind.
It is thus possible that, for some stars, stellar winds may
contribute to observational signatures that are typically
associated only with bipolar jets rooted in the accretion disk.

It is evident from Figure 5(a) that the pair of models for a
given star (i.e., the models for the HEG and HCGD values of the
observed parameters) can have drastically different predictions
for the stellar wind properties.
For many of the stars, the observational uncertainties drive the
uncertainties in the models.
Thus, it is worthwhile to ask {\em which} of the measured
parameters are most responsible for these large discrepancies.
Since the modeled $F_{\rm wind}$ is well correlated with
$F_{\rm wave}$, one can use
equations (\ref{eq:Pramdef})--(\ref{eq:rhoavdef}) in Section 3
to work out an approximate scaling relation for how
$F_{\rm wave}$ depends on the empirical input
parameters.\footnote{%
Doing this makes the implicit assumption that the accretion-driven
wave amplitude dominates the interior convection-driven wave
amplitude.  As in Cranmer (2008), the latter was assumed to remain
equal to the present-day Sun's value of 0.255 km s$^{-1}$.
Figure 3 shows that only a few of the cases modeled in this
paper have their {\em total} wave amplitude affected by this
convection-driven component.}
If one also assumes $F_{\rm wind} \propto F_{\rm wave}^{1.87}$
(see Figure 5(b)), then equation (\ref{eq:Fwind}) can be used
to estimate
\begin{equation}
  \dot{M}_{\rm wind} \propto \, \left(
  \frac{\dot{M}_{\rm acc}}{B_{\ast} \delta} \right)^{5.62}
  \frac{v_{c}^{9.37}}{M_{\ast} R_{\ast}^{8.24}} \,\, ,
\end{equation}
where it was found that the angular factor $\Delta\theta$ is roughly
proportional in magnitude to $\delta$, and small variations in the
stellar $\rho_{\ast}$ and $T_{\rm eff}$ are neglected.
The two most sensitive factors are the infalling clump speed
$v_c$ (which is related to the freefall speed $v_{\rm ff}$)
and the stellar radius $R_{\ast}$.
However, the accretion rate $\dot{M}_{\rm acc}$, the accretion
spot filling factor $\delta$, and the surface magnetic field
strength $B_{\ast}$ are also non-negligible contributors.
It is difficult to isolate just one or two of the observations
in, e.g., Table 1, that if improved would drastically improve
the model predictions.

It is also possible that the theoretical mass fluxes have been
underestimated systematically.
For example, the efficiency of conversion from the kinetic energy
of infalling clumps to MHD wave energy on the stellar surface
(equation \ref{eq:EA}) was computed by Scheurwater \& Kuijpers (1988)
in the ``far-field'' limit.
This approximation neglects energy near the base of the accretion
stream that is transient in nature, or is associated with possible
shocks, or is the result of center-to-edge inhomogeneities in the
streams (see, e.g., Romanova et al.\  2004).
The actual efficiency thus may be larger than was assumed in
equation (\ref{eq:EA}).
Also, if the stellar magnetic field has strong high-order multipole
components (i.e., departures from a dipole), the surface may be
peppered with a more evenly spread distribution of accretion spots
than is assumed here.
This would mean that the launching regions of the stellar wind may,
on average, be closer to the nearest accretion spots, and thus the
flux of MHD waves in these regions may be larger.

X-ray emission from the winds of T Tauri stars is generally
considered to be negligible when compared with that of the nearby
accretion shocks and closed-field coronal loops (see Figure 13
of Cranmer 2008).
This is similar to the case of dark ``coronal holes'' on the
surface of the Sun (e.g., Noci 1973; Zirker 1977; Cranmer 2009).
In open-field regions containing an accelerating stellar wind,
the density and pressure cannot build up to the higher levels
found in more confined and static regions.
The energy budget of the wind is thus dominated by the kinetic
energy lost in the outflow itself, and comparatively little is
left over for radiation at UV and X-ray wavelengths.
This paper thus does not compute the X-ray fluxes from these
polar open-field regions.

\section{Accretion Shocks}

The impact of magnetospheric accretion streams onto the stellar
surface generates standing shocks.
As the infall speeds are decelerated at the shock, both the
temperature and density increase.
There is also a postshock cooling zone, over which the density
continues to increase but the temperature decreases back down
to the undisturbed atmospheric value.
These regions have been suggested as potential sources of soft
X-rays for T Tauri stars (e.g., Lamzin 1999; Kastner et al.\  2002;
Grosso et al.\  2007; Robrade \& Schmitt 2007).
Although the surface areas of these spots are believed to be small
($\delta \sim 0.01$), the ballistic impact speeds are high enough
to generate temperatures of order 10$^{5}$--10$^{6}$ K.

The model of accretion shocks described here is a simplified
version of similar one-dimensional models in the literature
(Gullbring 1994; Calvet \& Gullbring 1998;
G\"{u}nther et al.\  2007; Sacco et al.\  2008).
Near the stellar surface, the magnetic field is assumed to be
oriented radially, and we ignore any structure or dynamics
transverse to the field.
The preshock velocity and density in the accretion streams are
given by $v_{1} = v_{c}$ and $\rho_{1} = \langle \rho \rangle$.
The immediate postshock conditions $v_{2}$ and $\rho_{2}$ are
computed from the standard Rankine-Hugoniot conditions
(Landau \& Lifshitz 1959) using an adiabatic ratio of specific
heats $\gamma = 5/3$.
The determination of the shock height, with respect to the
undisturbed stellar atmosphere, is described in Section 3.

Because the speed of infalling clumps $v_c$ is not necessarily
highly supersonic, it is necessary to use the complete version
of the shock-jump condition for the temperature,
\begin{equation}
  c_{2}^{2} \, = \,
  \frac{5 k_{\rm B} T_{\rm sh}}{3 \langle m \rangle} \, = \,
  \frac{(5 v_{1}^{2} - c_{1}^{2})(v_{1}^{2} + 3c_{1}^{2})}
  {16 v_{1}^2}
  \label{eq:c2sq}
\end{equation}
where $c_1$ is the sound speed in the accretion stream, $c_2$ is
the postshock sound speed, and $\langle m \rangle \approx
0.6 m_{\rm H}$ is the mean particle mass used here in order to
more accurately determine $T_{\rm sh}$.
We estimate $c_{1} = 13.8$ km s$^{-1}$, which is consistent with
a temperature in the accretion streams of $\sim$8000 K (see,
e.g., Martin 1996; Muzerolle et al.\  2001).
The commonly used approximate version of equation (\ref{eq:c2sq})
results when $c_{1} \rightarrow 0$.

Table 2 shows the postshock temperature $T_{\rm sh}$ and postshock
number density $n_{\rm post} = \rho_{2} / m_{\rm H}$ for the 28
empirical stellar cases studied here.
For these stars, the postshock temperatures range between about
0.02 and 4 MK.
There is a trend with age that can be described roughly by
\begin{equation}
  T_{\rm sh} \, \approx \, 5.2 \times 10^{5}
  \left( \frac{t}{1 \,\, \mbox{Myr}} \right)^{0.42} \,\, \mbox{K}
  \,\, ,  \label{eq:Tshfit}
\end{equation}
although there is substantial spread around this power-law
variation especially at the youngest ages.
It is important to note that the median value of $T_{\rm sh}$
is only 0.33 MK, and that there are only 3 out of 28 cases with
$T_{\rm sh} \geq 1$ MK.
This is lower than the typical values of 2--3 MK that
have been inferred from soft X-ray observations associated with
accretion shocks (e.g., Robrade \& Schmitt 2007).
It is possible that the reductions in the clump infall speed
below the ballistic free-fall speed ($v_{c} < v_{\rm ff}$)
described in Section 3 were overestimated.
However, had these deceleration factors been neglected, the
resulting models would have produced an unrealistically
{\em strong} level of X-ray emission that disagrees with the
observations (see Section 7).

The physical depth and structure of the postshock cooling zone
were computed using the analytic model of Feldmeier et al.\  (1997).
This model assumes that the radiative cooling rate
($Q_{\rm rad} = n_{e} n_{\rm H} \Lambda(T)$) exhibits a power-law
temperature dependence of $\Lambda \propto T^{-1/2}$.
The cooling zone has a vertical thickness given by equation (9)
of Feldmeier et al.\  (1997), which can be expressed as
\begin{equation}
  \Delta z \, = \, 0.407 \,
  \left( \frac{v_{2}}{50 \,\, \mbox{km} \,\, \mbox{s}^{-1}}
  \right)^{4}
  \left( \frac{\rho_{2}}{10^{-10} \,\, \mbox{g} \,\, \mbox{cm}^{-3}}
  \right)^{-1} \,\, \mbox{km} \,\, .
\end{equation}
For the T Tauri stars studied here, the thickness of the cooling
zone ranges from about 10 cm (for the youngest stars) to 100 km
(for the oldest stars).
The spatial dependence of density, temperature, and velocity in
the cooling zone is specified by equations (8)--(11) of 
Feldmeier et al.\  (1997), and the shapes of these curves
are reasonably close to those from numerical models that used
more recent calculations of the radiative cooling rates (e.g.,
Calvet \& Gullbring 1998; G\"{u}nther et al.\  2007).

\begin{figure}
\epsscale{1.17}
\plotone{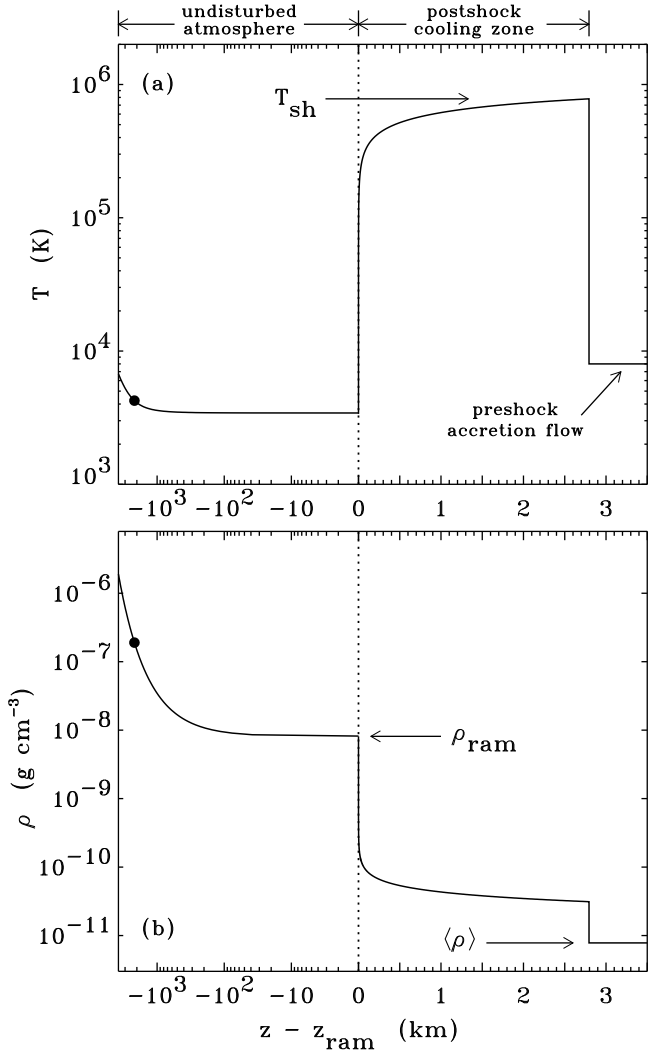}
\caption{Example accretion shock structure for the HCGD values
of AA Tau.  The temperature (a) and mass density (b) in the
vicinity of the shock and postshock cooling zone are shown.
The stellar photosphere is labeled by a filled circle, and
the base of the cooling zone is defined by $z_{\rm ram} = 0$
(dotted line).
The nonuniform abscissa coordinates are required to resolve
the various spatial scales in one plot.}
\end{figure}

Figure 6 illustrates how the temperature and density vary across
the shock, within the postshock cooling zone, and down to the
stellar photosphere.
The example shown here corresponds to the HCGD values of AA Tau.
The ``bottom'' of the cooling zone is defined as the height
$z_{\rm ram}$ at which the undisturbed atmosphere reaches a density
of $\rho_{\rm ram}$.
The shock itself sits at a height $\Delta z = 2.8$ km above the
bottom of the cooling zone, and the photosphere is about 2200 km
(or $\sim 3.1 H_{\ast}$) below it.

The predicted X-ray emission associated with the accretion shocks
is discussed in Section 7.
Note that the models presented here ignore any photoionization
heating of the {\em preshock} regions upstream of the impact site
(see, e.g., Calvet \& Gullbring 1998).
The temperatures that are reached in existing models of preshock
plasma appear to be low enough to not affect determinations of
the X-ray emission.
However, future models may need to include a self-consistent
treatment of photoionization in order to accurately predict the
UV emission from these regions.

\section{Closed Coronal Loops}

Young low-mass stars exhibit several observational signatures
of magnetic activity that are similar to the Sun's.
It has been suggested that the hottest components of observed
T Tauri X-rays ($T \approx 10$--30 MK) come from
stellar analogues of coronal loops and active regions
(e.g., Linsky 1985; Hartmann \& Noyes 1987; Preibisch 1997;
Stassun et al.\  2006, 2007; Robrade \& Schmitt 2007).
In this section, models of this kind of activity are described.
These models utilize recent developments in the theory of
turbulent coronal heating in closed-field regions.
The accretion-driven enhancements in photospheric MHD waves
(which are important for powering the polar winds) are assumed
to be responsible for increased energization at the footpoints
of the closed loops.

\subsection{Loop Geometry and Heating Rates}

The model of stellar coronal heating described below assumes
a time-steady distribution of loops without large-scale flows
along them.
In contrast, high-resolution solar observations show that the
closed-field corona is both highly time variable and full of
motion (e.g., Aschwanden 2006; Warren \& Winebarger 2007;
Patsourakos \& Klimchuk 2008).
The distribution of magnetic field strengths in the solar
corona may even be fractal in nature (Stenflo 2009), and the
observed loops are probably collections of many smaller
unresolved threads.
This paper does not attempt to simulate the full range of
microflaring and nanoflaring heating events that make up such
a dynamic corona.
Instead, the ``turbulence language'' used here is designed to
describe a statistical average over the large number of
small-scale impulsive heating events that must be present.

\begin{figure}
\epsscale{1.00}
\plotone{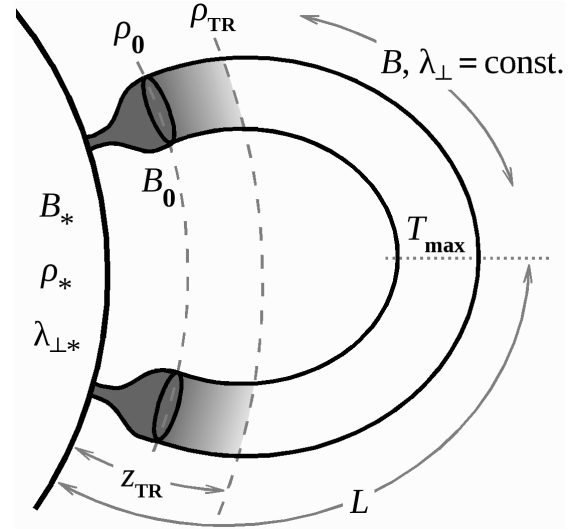}
\caption{Simplified portrayal of the assumed coronal loop
geometry.
Quantities at the photosphere are labeled with subscript
$\ast$.  Quantities at the flux-tube merging height have
subscript 0.  Quantities at the transition region have subscript TR.
For the T Tauri stars modeled in this paper the photosphere and
merging height are identical to one another, and thus the
``trumpet'' shaped regions between the photosphere and merging
height do not occur.
Magnetic field $B$ and turbulence correlation length
$\lambda_{\perp}$ are assumed to be constant everywhere above
the merging height.
The rapid drop in density from the merging height to the
TR is illustrated by a gradient in grayscale shading.}
\end{figure}

Figure 7 illustrates the adopted geometry for a representative
coronal loop and defines the plasma properties at several key
heights.
The photospheric values of the magnetic field $B_{\ast}$,
density $\rho_{\ast}$, and turbulence correlation length
$\lambda_{\perp \ast}$ (see below) are considered to be known
for each star.
In general, the photospheric magnetic field is fragmented into
small flux concentrations surrounded by regions of much weaker
field strength.
As height increases, the strong field in the concentrations
weakens and the cross-sectional area of the flux tube increases.
Eventually the flux tubes widen to the point of merging with their
neighbors and the magnetic field from the concentrations fills
the entire volume above that height.
For the Sun, the filling factor $f_{\ast}$ of strong
($\sim$ 1.5 kG) flux tubes in the photosphere ranges between
about 0.001 and 0.1, depending on the region.
The ``merging height'' where the flux tubes fill the volume
on the Sun tends to occur in the low chromosphere
(Cranmer \& van Ballegooijen 2005).
However, for T Tauri stars with saturated magnetic activity,
we will assume $f_{\ast} = 1$ (see Saar 2001), and thus the
merging height is assumed to be identical to the
photosphere.\footnote{%
It should be made clear that the filling factor $f_{\ast}$
applies only to the non-accreting regions of the stellar
surface.  The other filling factor discussed in this paper
(i.e., the surface fraction covered by accretion streams,
$\delta$) has no direct solar analogue.}

In Figure 7, the plasma properties at the merging height are
labeled with subscript 0.
Above the merging height, the closed magnetic flux tubes that fill
the corona are modeled as semicircular (i.e., half-torus) loops
with constant poloidal radius $r_{\perp}$ and half-length $L$.
The conservation of magnetic flux ($\nabla \cdot {\bf B} = 0$)
thus demands that the magnetic field strength $B_{0}$ is
constant along each loop.
Observed solar loops tend to have such a constant cross section,
but this may be related to their internal structure as collections
of thinner twisted strands (e.g., Klimchuk 2006).
The sharp transition region (TR) between the cool chromosphere
and the hot corona occurs above the merging height.
As far as the coronal loop models are concerned, the TR is
the effective lower boundary condition.
The temperature at the base of the TR is assumed to be a constant
value of $10^{4}$ K, and the density and pressure at that
location are outputs of the coronal heating model.

This paper uses a general phenomenological expression for the
coronal heating rate that was derived from analytic and
numerical studies of MHD turbulence.
The basic idea is that the photospheric waves (which are
produced by both internal convection and the accretion impacts)
cause a random shuffling of the magnetic field lines that thread
the surface.
Magnetic energy in the field is increased via the twisting,
shear, and braiding of the field lines (e.g., Parker 1972).
Treating this transport and dissipation of magnetic energy as
a type of ``turbulent cascade'' has been found to be useful
in parameterizing the eventual coronal heating
(van Ballegooijen 1986; Galsgaard \& Nordlund 1996;
G\'{o}mez et al.\  2000; Rappazzo et al.\  2008).
The expression for the volumetric heating rate is
\begin{equation}
  Q \, = \, \frac{\rho v_{\perp}^3}{\lambda_{\perp}}
  \left( \frac{\lambda_{\perp} V_{\rm A}}{v_{\perp} L}
  \right)^{\alpha}
  \label{eq:Qturb}
\end{equation}
where $\lambda_{\perp}$ is a transverse length scale that
represents an effective correlation length for the largest
eddies in the turbulent cascade (see also Hollweg 1986;
Chae et al.\  1998; Cranmer et al.\  2007).
The turbulent eddies are treated as Alfv\'{e}n waves that
counterpropagate along the loops with equal (i.e.,
``balanced'') amplitudes in the two directions along the field.
Specifically, the Alfv\'{e}n wave velocity amplitude $v_{\perp}$
and the correlation length $\lambda_{\perp}$ are taken at
the merging-height footpoints of the flux tubes.
The density $\rho$ and Alfv\'{e}n speed $V_{\rm A}$ vary with
position along the loops.

The magnitude of the Alfv\'{e}n wave activity on the surface must
be modified from the values given by equation (\ref{eq:vperpast}).
That expression gives the photospheric wave amplitude at the
poles, but here we need to evaluate the amplitude averaged over
the equatorial band of closed magnetic fields.
The same numerical code described in Section 4 was used to
compute a correction factor that takes account of this
geometrical effect.
The dimensionless $C_{\rm equ}$ factor is defined similarly to
$C_{\rm pol}$ (equation [\ref{eq:Cpol}]), but the integration
is taken from $\theta_{\rm in}$ to the equator ($\theta = \pi/2$),
and it is normalized by $\delta_{\rm equ}$.
Figure 4 shows how $C_{\rm equ}$ varies as a function of the
mean latitude of the accretion streams.
In contrast to $C_{\rm pol}$, which always is larger than one,
$C_{\rm equ}$ may be less than or greater than one.
When the accretion rings are near the poles, the wave power that
reaches the pole can be much larger than the average taken over
the (large) equatorial band.
However, when the accretion rings are near the equator, the wave
power that reaches the pole is significantly diminished below that
which remains near the equator.

Some additional comments on the form of equation (\ref{eq:Qturb})
are necessary.  The first part of the heating rate
($\rho v_{\perp}^{3} / \lambda_{\perp}$) is a cascade energy
flux that is identical to that derived by
von K\'{a}rm\'{a}n \& Howarth (1938) and Kolmogorov (1941)
for hydrodynamic turbulence.
There may be an order-unity constant that multiplies this
quantity (see, e.g., Hossain et al.\  1995; Pearson et al.\  2004),
but this is neglected here for simplicity.
The quantity in parentheses in equation (\ref{eq:Qturb}) is a
ratio of the cascade timescale $\tau_{c} = \lambda_{\perp}/v_{\perp}$
to the Alfv\'{e}n wave transit time along the loop
$\tau_{\rm A} = L/V_{\rm A}$.
This factor is a uniquely MHD phenomenon and does not appear
in hydrodynamic turbulence.
When the wave transit time is faster than the cascade time, the
nonlinear interactions between turbulent eddies become more
infrequent because ``wave packets'' do not spend as much
time interacting with one another (Kraichnan 1965).
Thus, the process of building up a power-law spectrum of
eddy scales is less efficient when $\tau_{\rm A} \ll \tau_{c}$.
In open flux tubes, this gives rise to an overall reduction of
the heating rate because the waves/eddies can escape easily.
However, in closed-field regions, the waves bounce back and
forth between the two ends of the loop and do not escape.
A lowering of the cascade efficiency leads to a slower
buildup of the turbulent magnetic energy.
This longer buildup eventually provides a {\em larger} output
of heat than would be the case with just the von K\'{a}rm\'{a}n
term (see, e.g., van Ballegooijen 1986).

The exponent $\alpha$ in equation (\ref{eq:Qturb}) describes
the effectively subdiffusive nature of the cascade in a coronal
plasma.
The wave-packet interactions discussed above are somewhat
muted by various MHD effects such as scale-dependent dynamic
alignment (Boldyrev 2005), and these nonlinearities modify the
efficiency of the coupling between eddies.
The case $\alpha = 0$ corresponds to pure hydrodynamic
turbulence.
Phenomenological expressions used for open field lines in the
solar wind (e.g., Oughton et al.\  2006; Cranmer et al.\  2007)
correspond to $\alpha = -1$ in the limit of
$\tau_{\rm A} \ll \tau_{c}$.
For coronal loops, various models of MHD turbulence have found
$\alpha = 1.5$ (G\'{o}mez et al.\  2000),
$\alpha = 2$ (van Ballegooijen 1986;
van Ballegooijen \& Cranmer 2008), as well as a continuum of
values between 1.5 and 2 that depends on the Alfv\'{e}n speed
and the wave amplitude (Rappazzo et al.\  2007, 2008).
In this paper we explore only this latter range of values
($1.5 \leq \alpha \leq 2$) as being appropriate for closed magnetic
loops energized by stellar turbulence.

\subsection{Results for Empirical Loop Distributions}

For each of the 28 stellar data points from Table 1, we created
a distribution of 100 loop lengths that spans the full range
of possible values for $L$.
The absolute minimum value $L_{\rm min}$ is defined geometrically
as the smallest possible toroidal radius at which the ``hole''
in the torus shrinks to zero size (i.e., $L_{\rm min} =
\pi r_{\perp}/2$).
To estimate the poloidal radius $r_{\perp}$, it is assumed that
both horizontal length scales in the photosphere ($r_{\perp}$
and $\lambda_{\perp}$) are proportional to the photospheric
scale height (see, e.g., Robinson et al.\  2004).
Estimated solar values for $r_{\perp} = 200$ km and
$\lambda_{\perp} = 75$ km (Cranmer et al.\  2007) are scaled
linearly with the ratio of the stellar to solar values of
$H_{\ast}$.
The maximum value for the loop length is assumed to reach up to
the inner edge of the accretion disk; i.e.,
$L_{\rm max} = r_{\rm in} - R_{\ast}$.

It should be noted that the precise definitions of the above limits
$L_{\rm min}$ and $L_{\rm max}$ are relatively unimportant.
We found that the smallest loops in this distribution are likely
to be ``submerged'' below the transition region, and thus they
would neither be heated to coronal temperatures nor contribute
to the X-ray emission.
Also, the largest loops are expected to appear so infrequently
that they should not contribute significantly to the disk-averaged
X-ray emission (see below).
Determining the effective lower limit length for which loops
are submerged in the chromosphere is done approximately by first
ignoring all loops with $L < z_{\rm TR}$.
The height of the transition region is estimated via hydrostatic
equilibrium,
\begin{equation}
  z_{\rm TR} \, \approx \, H_{\rm eff} \, \ln \left(
  \rho_{0} / \rho_{\rm TR} \right)  \,\, ,
\end{equation}
and the effective scale height $H_{\rm eff}$ takes account of
both gas pressure and wave pressure support:
$H_{\rm eff} = (c_{s}^{2} + v_{\perp \ast}^{2}) / \gamma g$.
For loops with $L > z_{\rm TR}$, we take account of partial
submerging by multiplying the expected X-ray fluxes by the
quantity $(1 - z_{\rm TR}/L)$.
For each stellar case in which the shortest loops are canceled out,
the minimum loop length $L_{\rm min}$ is redefined as $z_{\rm TR}$.

In some models of T Tauri star coronal heating (e.g.,
Ryan et al.\  2005; Jardine et al.\  2006),
the longest loops are often modeled as being unstable to
``breaking open'' into the stellar wind.
In many of these models, the gas pressure in loops increases
with increasing loop length $L$, and thus the longest loops
may have an internal gas pressure that exceeds the surrounding
magnetic pressure.
However, as described below, the use of equation (\ref{eq:Qturb})
for the heating rate gives rise to an {\em inverse} dependence
between loop gas pressure and loop length.
The shortest loops are thus the ones most liable to have gas
pressure exceed magnetic pressure, but in the present models
this does not occur.

The probability distribution $N(L)$ of loop lengths across the
stellar surface is assumed to be a power-law function of loop
length, with $N(L) \propto L^{-\varepsilon}$ and a sharp cutoff
below $L_{\rm min}$.
Various solar measurements have constrained the value of the
exponent $\varepsilon$ to be approximately 2.5
(Aschwanden et al.\  2000, 2008; Close et al.\  2003).
The models below use $\varepsilon = 2.5$ as a fiducial baseline,
but they also vary it as a free parameter to evaluate the
sensitivity to this assumption.
The distribution is normalized such that the equatorial band
of closed magnetic field is completely covered by loops.
The total surface area available for loops is assumed to be
given by $4\pi \delta_{\rm equ} R_{\ast}^{2}$.

For a power-law number distribution of loop lengths, the largest
loops (at the upper end of the distribution) do not contribute
strongly to any star-averaged quantities.
The {\em most probable} value of the loop length found on the
surface can be estimated from the first moment of the distribution,
\begin{equation}
  \langle L \rangle \, = \,
  \frac{\int dL \,\, N(L) \,\, L}{\int dL \,\, N(L)}
\end{equation}
and for a power-law distribution with a sharp cutoff below
$L_{\rm min}$, the most probable loop length is
$\langle L \rangle = L_{\rm min} (\varepsilon -1)/(\varepsilon -2)$.
For $\varepsilon = 2.5$, $\langle L \rangle = 3 L_{\rm min}$.
Since we find that $L_{\rm max}$ is typically a factor of
10$^2$--10$^4$ larger than $L_{\rm min}$, it is clear that the
shortest loops tend to dominate any observable star-averaged
properties of the corona.
Table 2 gives the most probable loop lengths for all 28
stellar cases.

For a given loop length $L$ and an assumed value of the $\alpha$
exponent in equation (\ref{eq:Qturb}), it is possible to solve for
the temperature and density along each loop.
The formulae of Martens (2008) were found to provide the most
general way of taking account of a heating rate that depends on
the spatially varying plasma properties.
Because $V_{\rm A} \propto \rho^{-1/2}$, there is an overall density
dependence of the heating rate, with $Q \propto \rho^{1 - \alpha/2}$.
When $\alpha = 2$ the heating rate is constant along the loop, but
when $\alpha < 2$ the heating is concentrated at the higher-density
footpoints.
Using the parameterization of Martens (2008), this proportionality
is described as
$Q \propto P^{1 - \alpha/2} / T^{1 - \alpha/2}$.
Equations (28)--(30) of Martens (2008) then provide solutions for
the peak temperature $T_{\rm max}$ and the pressure $P_{\rm TR}$
at the base of the transition region.
Neither the density $\rho_{\rm TR}$ nor the Alfv\'{e}n speed
$V_{\rm A, TR}$ at the base of the transition region are known at the
outset, but they need to be specified in equation (\ref{eq:Qturb})
in order to determine the heating rate.
Thus, the numerical code used to compute these properties iterates
to consistency on $\rho_{\rm TR}$ and $V_{\rm A, TR}$ from an
initial estimate.

Because of the large scale heights associated with a hot corona,
the loop models have been constructed under the assumption that
gas pressure remains constant along the length of each loop
(see, however, Serio et al.\  1981; Aschwanden \& Schrijver 2002).
Equation (25) of Martens (2008) is solved for $T(z)$ along the
loop, and the constraint of constant pressure provides the
density profile $\rho(z) = n(z) m_{\rm H}$.
Figure 8 shows an example of the temperatures and number densities
along a loop having the most probable length $\langle L \rangle$.
As in Figure 6, this plot shows the coronal properties for the
HCGD values of AA Tau.
The loop geometry has been ``unfolded'' in Figure 8 such that
the two photospheric endpoints occur at $z=0$ and $z=2L$.
All plasma quantities were computed between the transition region
height $z_{\rm TR}$ and the loop apex ($z=L$), and we assumed
symmetry such that both sides of the loop have equal properties.

\begin{figure}
\epsscale{1.17}
\plotone{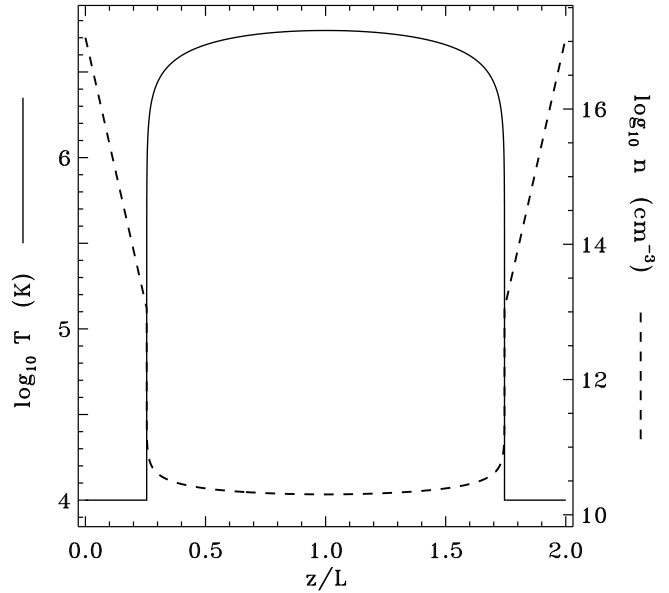}
\caption{Temperature (solid line) and number density (dashed
line) plotted as a function of position along a representative
coronal loop.
Distance is shown in units of the half-length $L$, which for
this example is the most-probable length
$\langle L \rangle = 18.6$ Mm for the HCGD values of AA Tau.
This model utilized the iterated coronal heating exponent
$\alpha_{\rm R} = 1.98$ (see Table 2).}
\end{figure}

In addition to computing model coronal loops for constant values
of $\alpha$, another set of models was constructed with a
parameter-dependent choice for $\alpha$ inspired by the results
of Rappazzo et al.\  (2008).
In a set of numerical MHD turbulence simulations for solar
coronal loops, Rappazzo et al.\  (2008) found that $\alpha$
should depend on the ratio $\cal R$ of the footpoint turbulence
amplitude to the coronal Alfv\'{e}n velocity.
Using the definition of the $\alpha$ exponent from equation
(\ref{eq:Qturb}), their numerical results have been fit with the
following approximate relation,
\begin{equation}
  \alpha_{\rm R} \, \approx \,
  \frac{2 + 420 {\cal R}}{1 + 280 {\cal R}}
  \label{eq:alrapp}
\end{equation}
where ${\cal R} = v_{\perp 0} / V_{\rm A, TR}$ and we use the
variable name $\alpha_{\rm R}$ to refer specifically to the
iterated\footnote{%
Note that equation (\ref{eq:alrapp}) depends on $V_{\rm A, TR}$,
which is known for a given loop model only after the iteration
process described above.  The loop models constructed with
a variable $\alpha_{\rm R}$ thus involved an additional round of
iteration in order to find the most self-consistent parameters.}
model value given by equation (\ref{eq:alrapp}).
It is important to note that the exponent $\alpha$ that is
defined here differs from the similarly named exponent used by
Rappazzo et al.\  (2008).
If the latter is renamed $\beta$, then
$\alpha = (2\beta + 3)/(\beta + 2)$, and the direct fit of
the Rappazzo et al.\  (2008) results yielded
$\beta_{\rm R} \approx (140 {\cal R})^{-1}$.
In any case, equation (\ref{eq:alrapp}) spans a range of
possible values between 1.5 and 2.
In the loop models computed for this paper, the ratio
${\cal R}$ spans several orders of magnitude between about
10$^{-5}$ and 10$^{-2}$.
Since there are 28 stellar cases and 100 possible loop lengths
for each case, there were eventually 2800 individual values of
the ``preferred'' $\alpha_{\rm R}$ exponent.
For this large set of values, the mean value of $\alpha_{\rm R}$
was found to be 1.868, with a standard deviation of 0.101 about
that mean.
For a given star, as $L$ increases so does $\alpha_{\rm R}$.

Table 2 shows example solutions for $T_{\rm max}$, $\alpha_{\rm R}$,
the basal number density $n_{\rm TR} = \rho_{\rm TR} / m_{\rm H}$,
and the number density at the peak of the loop $n_{\rm top}$.
These models are shown for the preferred case of
$\alpha = \alpha_{\rm R}$ and for the most probable loop lengths
$\langle L \rangle$ as defined above.
It is evident that the loop heating rate given by equation
(\ref{eq:Qturb}) provides hot coronal emission with peak
temperatures of order 5--10 MK at densities around
10$^{10}$ cm$^{-3}$.
Figure 9 also shows how $T_{\rm max}$ and $n_{\rm top}$ vary as
a function of loop length and $\alpha$ for an example stellar case
(the HEG values of GM Aur).
There is generally more coronal heating for larger values of
the $\alpha$ exponent.

\begin{figure}
\epsscale{1.17}
\plotone{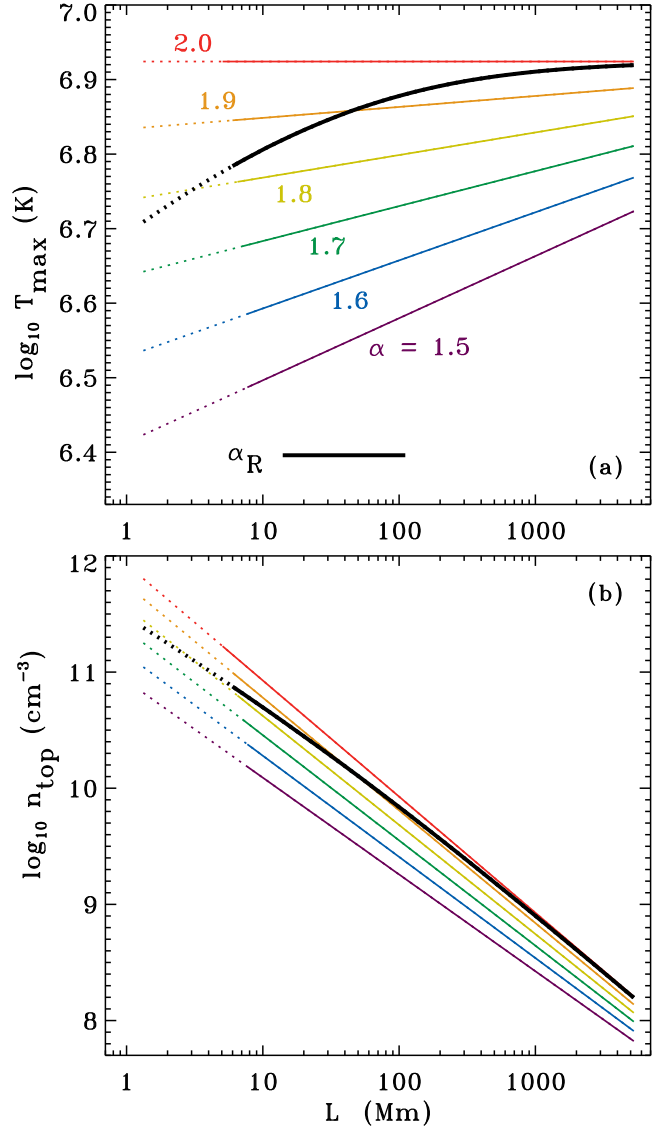}
\caption{Parametric plots of (a) loop-top temperature
$T_{\rm max}$ and (b) loop-top number density $n_{\rm top}$
versus loop length $L$.
Results are shown for a range of possible values of the
coronal heating exponent $\alpha$ for an example stellar
case (the HCGD values for GM Aur).
Colors and labels show results computed for specific constant
values of $\alpha$, and the thick black curve denotes models
created with the iterated values of $\alpha_{\rm R}$ from
equation (\ref{eq:alrapp}).
Dotted portions of each curve denote loops submerged below
the TR, which are assumed not to contribute to X-ray emission.}
\end{figure}

The variation of the coronal heating properties with loop length
$L$ seen in Figure 9 can be understood from the basic scaling
relations of Martens (2008).
Once the parameter dependence of the heating rate is taken into
account, the loop-top temperature varies as
$T_{\rm max} \propto L^{(2 - \alpha)/(3 + 2\alpha)}$.
This exponent varies from 1/12 (for $\alpha = 1.5$) to zero
(for $\alpha = 2$).
This scaling is less sensitive to loop length than the classical
proportionality derived by Rosner et al.\  (1978) for a constant
heating rate ($T_{\rm max} \propto L^{4/7} Q^{2/7}$).
The basal loop pressure scales as
$P_{\rm TR} \propto L^{(3 - 5\alpha)/(3 + 2\alpha)}$, and the
exponent varies between --3/4 and --1 for the allowed range of
$\alpha$ values.
In this case, the scaling with loop length is in the opposite
sense as Rosner et al.\  (1978), who found that
$P_{\rm TR} \propto L^{5/7} Q^{6/7}$ for a constant $Q$.
For the turbulent heating models of this paper, the fact that
$P_{\rm TR}$ decreases with increasing $L$ means that the shortest
loops are the ones most in danger of having their gas pressure
exceed their magnetic pressure.
For the 28 stellar cases, though, the ratio of gas pressure to
magnetic pressure (for the shortest coronal loops in each case)
ranges from about $10^{-4}$ to $10^{-2}$.
Thus, the equatorial coronal loops modeled here do not appear
to be unstable to breaking open into a stellar wind.

Finally, it is useful to confirm the validity of the adopted
heating rate and loop modeling procedure by applying
representative values for well-observed solar loops.
The Sun's photospheric filling factor for strong-field magnetic
flux tubes is much smaller than unity.
Here we assume $f_{\ast} = 0.05$, which is consistent with a
mean field strength at the merging height of about 70 G.
This corresponds to a weak active region or small sunspot.
At the merging height, the transverse footpoint velocity
amplitude $v_{\perp} \approx 1$ km s$^{-1}$, and the correlation
length $\lambda_{\perp} \approx 350$ km.
Taking the typical value $\alpha = 1.9$ for a representative
loop with $L = 50$ Mm, the iteration process described above
gives rise to realistic values for the basal Alfv\'{e}n speed
$V_{\rm A, TR} \approx 900$ km s$^{-1}$, the transition region
number density $n_{\rm TR} \approx 3 \times 10^{10}$ cm$^{-3}$,
and the peak temperature $T_{\rm max} \approx 1.2$ MK.
The temperature is slightly on the low side for an active region,
but it is consistent with much of the fine loop structure
observed in the extreme UV by {\em TRACE} (Lenz et al.\  1999;
Aschwanden 2006).
Applying a loop length distribution with $\varepsilon = 2.5$
and computing the surface average of the energy flux
($F \sim QL$), this solar model gives a mean value of
$5 \times 10^{6}$ erg cm$^{-2}$ s$^{-1}$.
This is in the middle of the empirically allowed range for
active region loops (e.g., Withbroe \& Noyes 1977).

\section{X-Ray Luminosities}

The accretion shocks (Section 5) and coronal loops (Section 6)
are both expected to give rise to significant X-ray emission.
Their total X-ray luminosities $L_{\rm X}$ are estimated here
and compared with existing observations.
The calculations of X-ray flux are done under the optically thin
assumption that all radiation escapes from the emitting regions.
The absorption by intervening material is not taken into account
because that has been shown to be sensitively dependent on both
the assumed magnetic geometry and the observer's orientation
(e.g., Gregory et al.\  2007).
Future work should include these effects and compute time-dependent
X-ray light curves for observers at a range of inclinations.

The emissivity of each volume element of plasma was treated using
version 5.2.1 of the CHIANTI atomic database (Dere et al.\  1997;
Landi et al.\  2006).
The models used a standard solar abundance mixture
(Grevesse \& Sauval 1998) and collisional ionization balance
(Mazzotta et al.\  1998) to compute the emissivity as a function
of wavelength over the X-ray bandpasses of interest.
This provided a bandpass-dependent radiative loss rate
($Q_{\rm X} = n_{e}^{2} \Lambda_{\rm X}(T)$) that was then used
to sum up the contributions of the emitting volume elements.
For structures near the surfaces of the stars, it is valid to
estimate the X-ray luminosity as
\begin{equation}
  L_{\rm X} \, = \, 4\pi \delta R_{\ast}^{2}
  \int dz \,\, [n_{e}(z)]^{2} \Lambda_{\rm X}[T(z)]  \,\, ,
\end{equation}
where the fraction $\delta$ is used for the accretion shock
regions and it is replaced by $\delta_{\rm equ}$ for the equatorial
coronal loop regions.

\begin{figure}
\epsscale{1.17}
\plotone{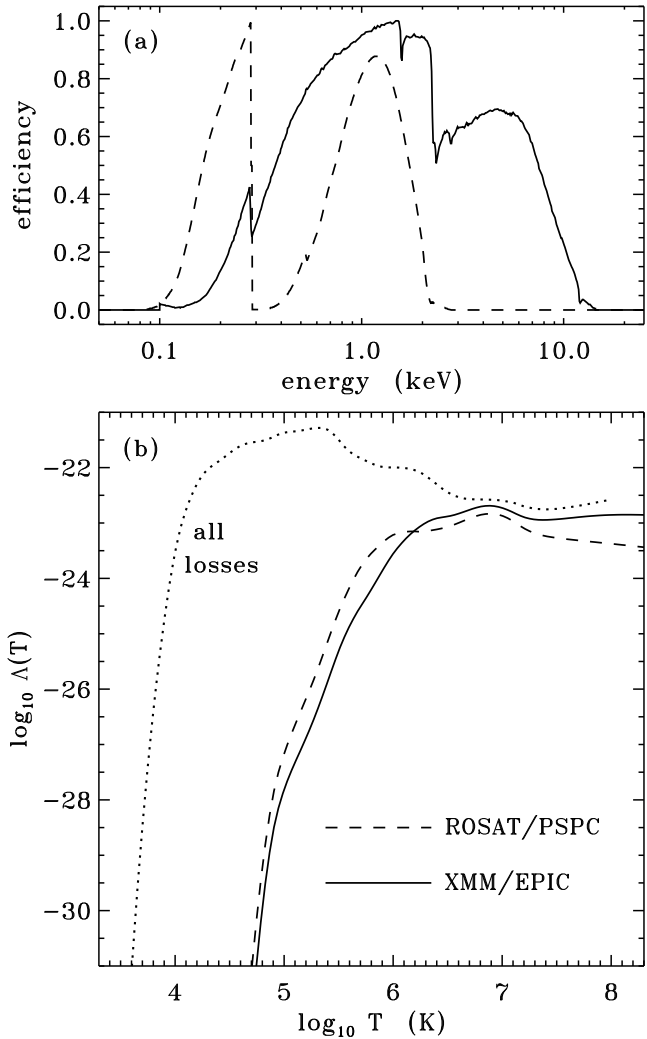}
\caption{(a) Normalized X-ray efficiencies for the {\em ROSAT}
PSPC (dashed line) and for the medium thickness filter on
{\em XMM-Newton's} EPIC-pn camera (solid line).
(b) Temperature dependence of optically thin radiative
loss functions computed for the {\em ROSAT} bandpass (dashed
line), the {\em XMM} bandpass (solid line), and for all photons
(dotted line).}
\end{figure}

Figure 10 shows the relevant X-ray response functions and their
associated radiative loss rates.
Two observational bandpasses are discussed in this paper.
The first is that of the {\em ROSAT} Position Sensitive
Proportional Counter (PSPC), with the X-ray efficiency function
specified by Judge et al.\  (2003).
The sensitivity is nonzero between about 0.1 and 2.4 keV, with
a minimum around 0.3 keV that separates the hard and soft bands.
The second bandpass shown is that of the {\em XMM-Newton}
European Photon Imaging Camera (EPIC), and Figure 10(a) shows
the sensitivity function for the medium thickness filter on the
EPIC-pn camera (e.g., Dahlem \& Schartel 1999;
Str\"{u}der et al.\  2001).
Figure 10(b) shows the result of folding the optically thin
emissivities computed with CHIANTI through the two response
functions, as well as the full radiative loss curve that
includes photons of all wavelengths (see, e.g.,
Raymond et al.\  1976; Schmutzler \& Tscharnuter 1993).

\begin{deluxetable*}{cccccccccc}
\tablecaption{Observed and Modeled X-ray Luminosities\tablenotemark{a}}
\tablewidth{0pt}

\tablehead{
\colhead{} &
\multicolumn{2}{c}{Observed} &
\colhead{} &
\multicolumn{6}{c}{Modeled} \\
\cline{2-3}
\cline{5-10}
\colhead{} &
\colhead{} &
\colhead{} &
\colhead{} &
\colhead{} &
\colhead{} &
\colhead{} &
\colhead{} &
\colhead{} &
\colhead{} \\
\colhead{} &
\colhead{} &
\colhead{} &
\colhead{} &
\colhead{Shocks} &
\colhead{Shocks} &
\colhead{Loops, $\alpha = 1.5$} &
\colhead{Loops, $\alpha = 2$} &
\colhead{Loops, $\alpha = \alpha_{\rm R}$} &
\colhead{Loops, $\alpha = \alpha_{\rm R}$} \\
\colhead{Object} &
\colhead{\em{ROSAT}} &
\colhead{\em{XMM}} &
\colhead{} &
\colhead{(\em{ROSAT})} &
\colhead{(\em{XMM})} &
\colhead{(\em{ROSAT})} &
\colhead{(\em{ROSAT})} &
\colhead{(\em{ROSAT})} &
\colhead{(\em{XMM})}
}

\startdata

 AA Tau &
  29.642 &  30.094 & &  26.775 &  26.094 &
   29.071 &   30.777 &   29.871 &   30.013 \\
 &
  29.642 &  30.094 & &  29.579 &  28.962 &
   26.231 &   29.254 &   29.100 &   29.249 \\
 BP Tau &
  29.849 &  30.135 & &  25.488 &  24.838 &
   26.031 &   28.344 &   27.999 &   28.132 \\
 &
  29.849 &  30.135 & &  29.465 &  28.760 &
   25.615 &   28.211 &   28.000 &   28.144 \\
 CY Tau &
  $<$29.797 &  29.288 & &  30.693 &  30.462 &
   27.857 &   29.833 &   29.120 &   29.266 \\
 &
  $<$29.797 &  29.288 & &  30.257 &  29.750 &
   25.972 &   28.710 &   28.513 &   28.663 \\
 DE Tau &
  $<$29.458 &  \ldots & &  26.936 &  26.265 &
   29.856 &   31.672 &   30.734 &   30.908 \\
 &
  $<$29.458 &  \ldots & &  28.966 &  28.262 &
   27.498 &   30.444 &   30.218 &   30.371 \\
 DF Tau &
  29.810 &  \ldots & &  28.096 &  27.405 &
   28.784 &   30.958 &   30.248 &   30.396 \\
 &
  29.810 &  \ldots & &  29.243 &  28.532 &
   28.539 &   30.814 &   30.157 &   30.302 \\
 DK Tau &
  $<$29.348 &  29.962 & &  \ldots\tablenotemark{b} &
                           \ldots\tablenotemark{b} &
   28.407 &   30.458 &   29.689 &   29.835 \\
 &
  $<$29.348 &  29.962 & &  25.635 &  24.982 &
   25.960 &   28.995 &   28.863 &   29.011 \\
 DN Tau &
  29.751 &  30.063 & &  27.942 &  27.221 &
   28.637 &   30.447 &   29.606 &   29.750 \\
 &
  29.751 &  30.063 & &  28.098 &  27.395 &
   27.205 &   29.860 &   29.505 &   29.652 \\
 DO Tau &
  $<$29.316 &  \ldots & &  26.033 &  25.368 &
   28.119 &   30.093 &   29.348 &   29.493 \\
 &
  $<$29.316 &  \ldots & &  30.596 &  29.889 &
   27.562 &   29.706 &   29.050 &   29.195 \\
 DS Tau &
  $<$29.877 &  \ldots & &  23.213 &  22.477 &
   29.394 &   31.136 &   30.248 &   30.393 \\
 &
  $<$29.877 &  \ldots & &  30.394 &  29.852 &
   28.306 &   30.449 &   29.612 &   29.766 \\
 GG Tau &
  $<$29.029 &  \ldots & &  25.730 &  25.081 &
   29.060 &   30.629 &   29.650 &   29.798 \\
 &
  $<$29.029 &  \ldots & &  30.075 &  29.405 &
   26.708 &   29.313 &   28.956 &   29.099 \\
 GI Tau &
  29.382 &  29.921 & &  26.491 &  25.827 &
   28.820 &   30.641 &   29.814 &   29.958 \\
 &
  29.382 &  29.921 & &  29.523 &  28.819 &
   28.597 &   30.682 &   29.976 &   30.120 \\
 GM Aur &
  29.750 &  \ldots & &  17.593 &  16.452 &
   28.633 &   30.490 &   29.609 &   29.754 \\
 &
  29.750 &  \ldots & &  24.557 &  23.908 &
   28.518 &   30.500 &   29.674 &   29.822 \\
 HN Tau &
  $<$29.709 &  \ldots & &  31.208 &  30.986 &
   28.751 &   30.609 &   29.605 &   29.748 \\
 &
  $<$29.709 &  \ldots & &  30.790 &  30.921 &
   28.757 &   30.626 &   29.673 &   29.816 \\
 UY Aur &
  $<$29.403 &  \ldots & &  25.957 &  25.307 &
   28.119 &   30.323 &   29.524 &   29.669 \\
 &
  $<$29.403 &  \ldots & &  29.688 &  28.985 &
   27.825 &   30.097 &   29.461 &   29.608 \\

\enddata
\tablenotetext{a}{All X-ray luminosities are listed as
base-10 logarithms of $L_{\rm X}$ in units of erg s$^{-1}$.}
\tablenotetext{b}{Postshock temperature $T_{\rm sh}$ was
too low to generate significant X-ray emission; see Figure 10.}

\end{deluxetable*}

Table 3 lists the observed and predicted X-ray luminosities for
a variety of cases.
The {\em ROSAT} observations were taken from
Neuh\"{a}user et al.\  (1995).
Two of the measurements listed under the {\em ROSAT} column
(for DF Tau and GM Aur) were made with the {\em Einstein}
Imaging Proportional Counter (IPC) which has a similar X-ray
bandpass as the {\em ROSAT} PSPC (Damiani et al.\  1995).
The {\em XMM-Newton} measurements were reported by
G\"{u}del et al.\  (2007a).
For the 4 out of 14 stars for which there are firm observations
from both telescopes, the ratios of {\em XMM} to {\em ROSAT}
luminosities range between 1.9 and 3.5.
The curves in Figure 10(b) show a similar range of enhancement
for temperatures between about 20 and 120 MK.

The representative example of a solar coronal loop discussed at
the end of Section 6 can be used to help confirm the validity
of the theoretical X-ray luminosity estimates.
For this example, using the {\em ROSAT} bandpass and integrating
over the full loop-length distribution yielded
$L_{\rm X} = 1.5 \times 10^{27}$ erg s$^{-1}$, or a ratio
$L_{\rm X} / L_{\odot} = 4 \times 10^{-7}$.
This is inside the observed range of variability for the Sun
(Judge et al.\  2003).

\begin{figure}
\epsscale{1.17}
\plotone{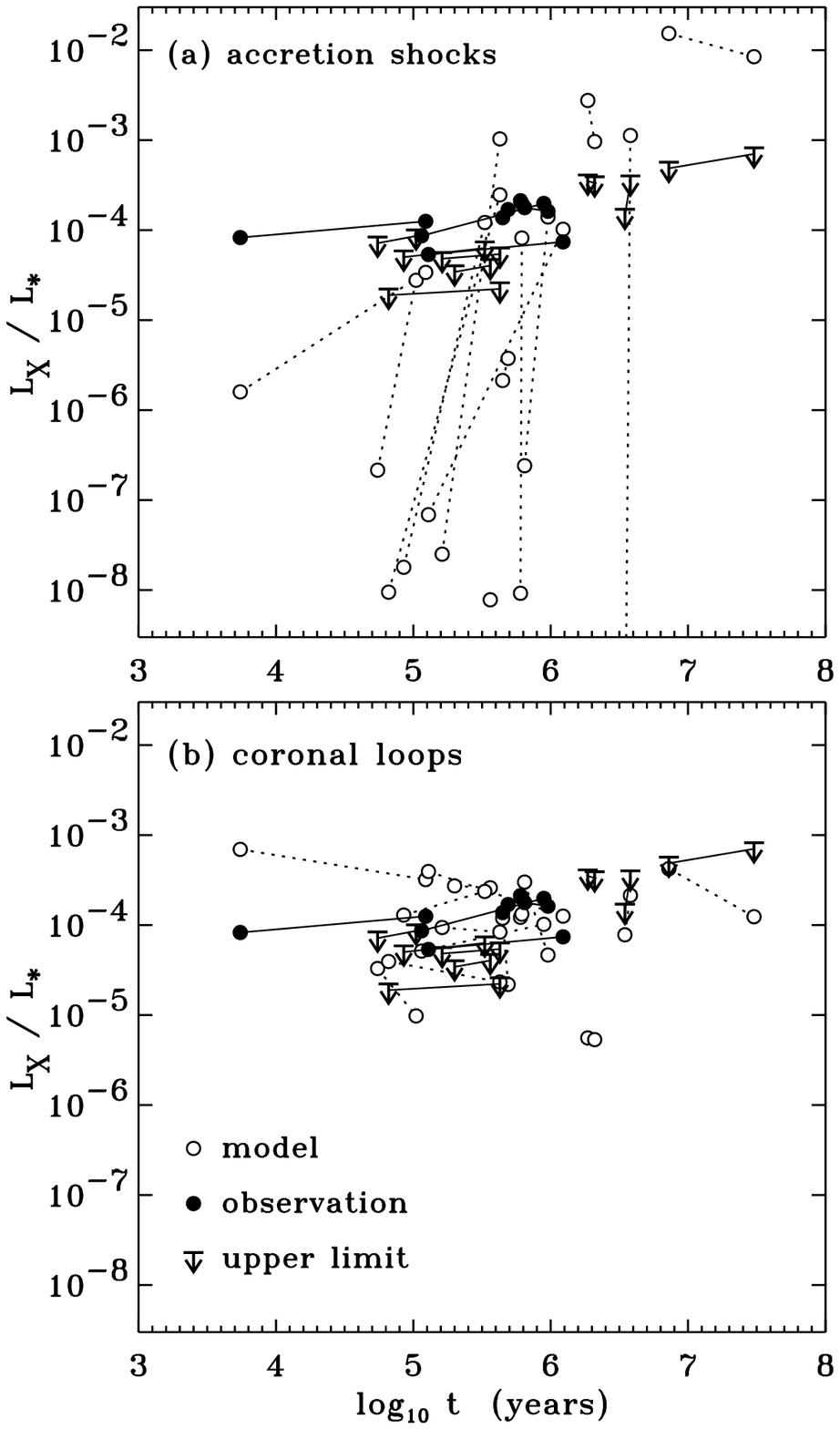}
\caption{Individual results for $L_{\rm X}/L_{\ast}$ shown as a
function of age, corresponding to (a) the accretion shock models
and (b) coronal loop models computed with the optimized
$\alpha = \alpha_{\rm R}$ exponents.
Model values computed with the {\em ROSAT} bandpass (open circles)
are compared against direct {\em ROSAT} observations (filled circles)
and upper limits (downward arrows).}
\end{figure}

Figure 11 shows the ratios $L_{\rm X} / L_{\ast}$ for the 28
stellar cases, computed using the {\em ROSAT} bandpass for both
the accretion shock models and a representative set of coronal
loop models (i.e., the optimized case of $\alpha = \alpha_{\rm R}$).
Table 3 also lists the predicted $L_{\rm X}$ values for these models,
as well as the corresponding {\em XMM} bandpass luminosities.
Loop model results are also given (for the {\em ROSAT} bandpass
only) for the cases of fixed values of $\alpha = 1.5$ and 2.
Figure 11(a) shows the modeled X-ray luminosities for the
accretion shock models.
These values span a much wider range than the observations.
The youngest stars that generally have the highest accretion
rates (but lowest values of $v_{c}$ and $T_{\rm sh}$) tend to
have negligibly small predicted values of $L_{\rm X}$ from
the accretion shocks.
Figure 11(b) shows the predicted X-ray luminosities from the
coronal loops, and these appear to fall more closely within
the observed range of values.

Note from Table 3 that the modeled {\em XMM} luminosities for
the accretion shock models are typically lower than the
{\em ROSAT} bandpass values.
Apart from one outlier (the HCGD value of HN Tau), nearly all
of the modeled stellar cases have ratios of {\em XMM} to
{\em ROSAT} $L_{\rm X}$ between 0.07 and 0.6.
For the coronal loop models, however, this ratio spans a higher
range between about 0.6 and 1.7.
The values less than 1 apply only for models that use
$\alpha \lesssim 1.6$, and for the optimized models that use
$\alpha = \alpha_{\rm R}$ the range of {\em XMM} to {\em ROSAT}
ratios is 1.3 to 1.5.
This is consistent with the fact that most of the loop-top
temperatures $T_{\rm max}$ are between about 5 and 10 MK
(see Figure 10(b)).
In any case, these ratios are well below the observed range
of 1.9 to 3.5.
This may indicate that the present models underestimate the
coronal temperatures.

\begin{figure}
\epsscale{1.17}
\plotone{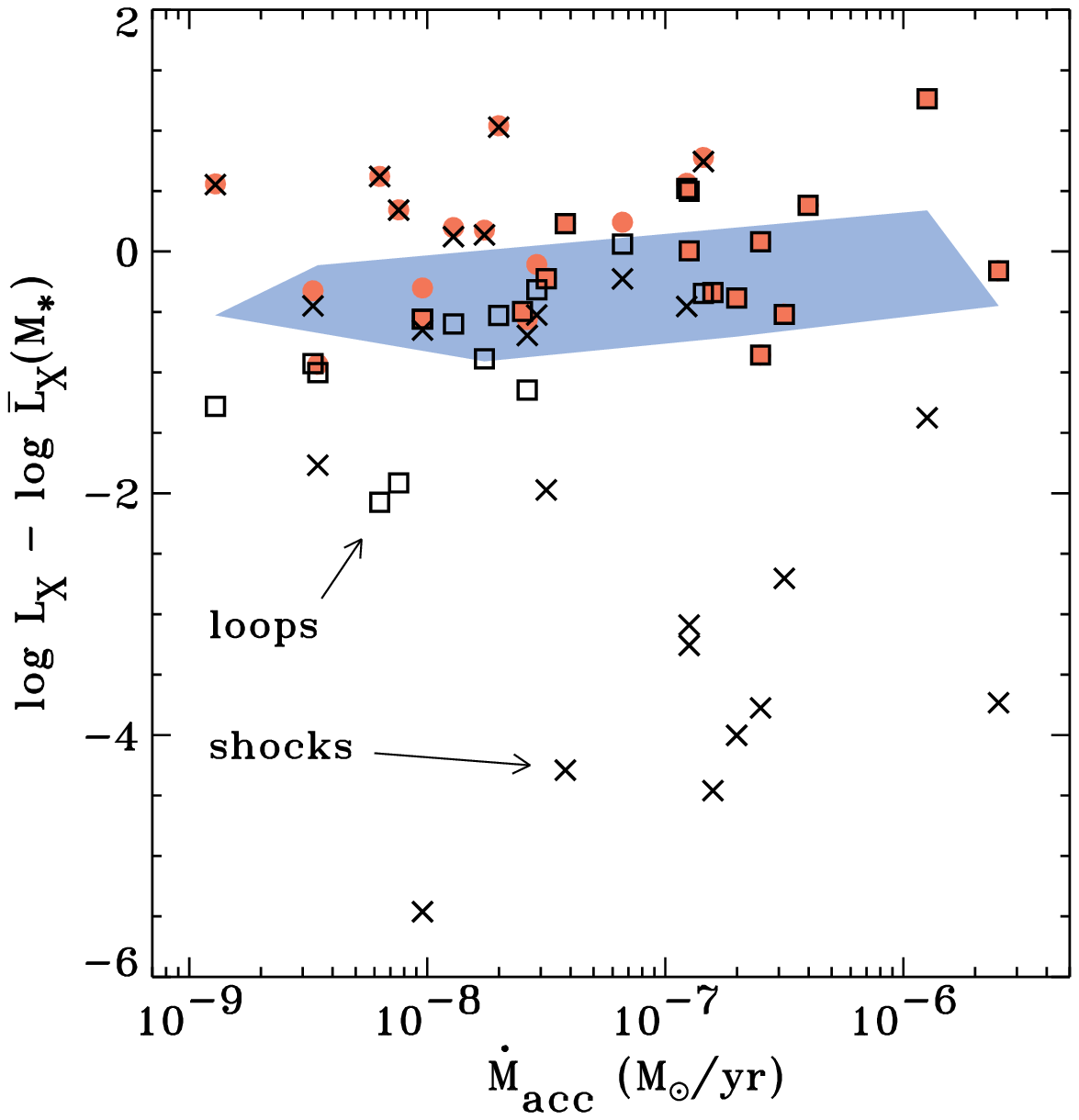}
\caption{Residual X-ray luminosity (i.e., deviation of $L_{\rm X}$
from the mean observational trend with stellar mass
$\bar{L}_{\rm X} (M_{\ast})$) shown as a function of mass
accretion rate.
Model results include individual luminosities from the accretion
shocks (crosses), the coronal loop models computed with optimized
$\alpha_{\rm R}$ (squares), and the sum of these two components
(red filled circles).
The approximate region subtended by the {\em ROSAT} observations
is illustrated by a blue polygon.}
\end{figure}

In order to assess the {\em relative contributions} from accretion
shocks and coronal loops, Figure 12 shows the modeled X-ray
luminosities (again just using the {\em ROSAT} bandpass) as a
function of the stellar accretion rate.
In this plot, we follow Drake et al.\  (2009) and plot the
residual X-ray luminosity, in which the general trend for
higher-mass stars to have a larger $L_{\rm X}$ has been first
removed.
The mean mass dependence has been taken from the {\em Chandra}
COUP data sample (Preibisch et al.\  2005), with
\begin{equation}
  \log \bar{L}_{\rm X} (M_{\ast}) \, = \,
  30.34 + 1.13 \log (M_{\ast}/M_{\odot}) \,\, .
\end{equation}
Figure 12 shows the individual $L_{\rm X}$ contributions from
accretion shocks and coronal loops for each stellar case, as
well as their sum.
The totals generally fall within the range of observed values
(shown as an irregular blue polygon).
Of the 28 stellar cases shown in Figure 12, 8 of them have X-ray
luminosities that are clearly dominated by the accretion shocks
(i.e., they have more than a 75\% relative contribution from the
shock emission) and 16 are clearly dominated by the coronal loops
(with more than a 75\% contribution from the loops).
The 4 remaining cases have a rough balance between the two
components; these correspond to the HCGD values of
BP Tau, DE Tau, GI Tau, and UY Aur.
Figure 12 shows that, generally, the highest values of 
$\dot{M}_{\rm acc}$ correspond to the cases dominated by
coronal loops.
The stars with the lowest accretion rates are the ones with
$L_{\rm X}$ dominated by the accretion shocks.

The recent study of Drake et al.\  (2009) indicated the existence
of an anticorrelation between accretion rate and X-ray luminosity
for some T Tauri stars.
This appears to be consistent with the well-known trend for
WTTS (weak-lined T Tauri stars, which are not accreting) to be
more X-ray bright than accreting CTTS (e.g.,
Neuh\"{a}user et al.\  1995; Flaccomio et al.\  2003a,b;
Telleschi et al.\  2007; Briggs et al.\  2007).
However, neither the observational data nor the model results
shown in Figure 12 yield such an anticorrelation.
If one examines just the modeled coronal loop $L_{\rm X}$ values,
it is clear that there tends to be a {\em positive} correlation
with accretion rate.
This makes sense since the coronal heating is largely driven
by MHD turbulence excited in the accretion streams.
The X-ray luminosities associated with the accretion shocks
are roughly anticorrelated with accretion rate.
When the two components are summed, however, these effects
largely cancel each other out, and the result is little to no
correlation with accretion rate.
Of course, the present sample does not span the wider range of
ages and accretion rates that one would find when comparing
WTTS and CTTS.

Another way to assess the validity of the models is to perform
a direct star-by-star $\chi^{2}$ analysis of the differences
between the observed and modeled X-ray luminosities.
This analysis uses the summed $L_{\rm X}$ from both the accretion
shock and coronal loop regions for each stellar case.
Also, only the most definitive observations are used.
Only the 6 {\em ROSAT} observations with firm measurements
(e.g., AA Tau, BP Tau, DF Tau, DN Tau, GI Tau, and GM Aur)
and the 2 {\em XMM} measurements that exist for stars having
only {\em ROSAT} upper limits (e.g., CY Tau and DK Tau) are
used in this analysis.
This results in 8 stars (or 16 unique model data points) to
compare with the observations.
The $\chi^{2}$ statistic computed for a given set of models
is defined as
\begin{equation}
  \chi^{2} \, = \, \frac{1}{\sigma^2} \sum \left(
  \log L_{\rm X, obs} - \log L_{\rm X, model} \right)^{2} 
  \,\, ,
  \label{eq:chi2}
\end{equation}
where $\sigma = 0.5$ dex was adopted as the logarithm of a
representative observational uncertainty (see, e.g.,
Neuh\"{a}user et al.\  1995) and the above sum is taken
over the 16 data points.
Each term in the sum uses the appropriate bandpass for
each of the $L_{\rm X}$ values; i.e., each term compares
either {\em ROSAT} observations to {\em ROSAT}-bandpass models,
or {\em XMM} observations to {\em XMM}-bandpass models.

The $\chi^{2}$ quantity defined in equation (\ref{eq:chi2})
constrains the probability that the observed and modeled data
sets are in agreement with one another.
Assuming normally distributed uncertainties, this probability
is given by Press et al.\  (1992) as
\begin{equation}
  P \, \equiv \, Q (\chi^{2} | \nu)
  = \frac{1}{\Gamma (\nu / 2)}
  \int_{\chi^{2} / 2}^{\infty} e^{-t} t^{(\nu / 2)-1} dt
  \,\, ,
\end{equation}
where $\nu$ is the effective number of degrees of freedom,
which we take as 16 (the number of data points), and
$\Gamma(x)$ is the complete Gamma function.
When $\chi^{2} \ll \nu$ the above probability
approaches unity (i.e., the modeled luminosities are a good match
to the observed luminosities), and when $\chi^{2} \gg \nu$
the above probability is negligibly small.

\begin{figure}
\epsscale{1.17}
\plotone{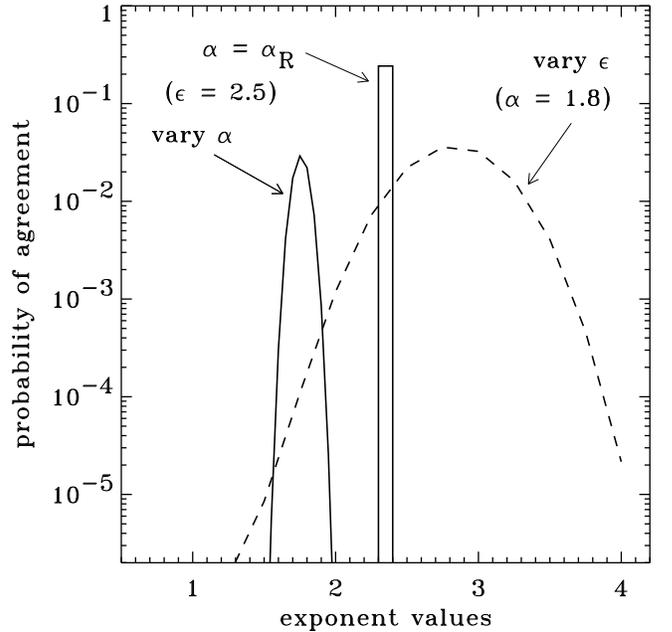}
\caption{Computed probabilities of agreement between X-ray
observations and the modeled total (shock plus loop) values
of $L_{\rm X}$ for a set of 8 well-observed stars.
Models are shown that vary the coronal heating exponent $\alpha$
(solid curves) and those that vary the loop number distribution
exponent $\varepsilon$ (dashed curve).
For clarity, the single histogram corresponding to the optimized
$\alpha_{\rm R}$ models is placed to the right of the other
models that assume a range of constant $\alpha$ exponents (i.e.,
the plotted abscissa value for this case was chosen arbitrarily).}
\end{figure}

Figure 13 shows the results of several calculations of the
above probability.
Because there were no real free parameters to vary for
the accretion shock models, all of the models shown in
Figure 13 vary either the coronal-loop heating exponent
$\alpha$ or the loop number distribution exponent
$\varepsilon$.
When $\alpha$ was varied (keeping $\varepsilon$ fixed at its
standard value of 2.5), the highest probability of agreement
with the observations was clearly found for the optimized case
of $\alpha = \alpha_{\rm R}$.
This model resulted in a reduced $\chi^{2}/\nu = 1.22$, which
corresponds to a probability of 24\%.
For fixed values of $\alpha$, the exponent was varied between
1.5 and 2 with an increment of 0.05.
The highest-probability model of these was $\alpha = 1.75$,
with a reduced $\chi^{2} / \nu$ of 1.77 and a probability
of about 3\%.

Varying the loop-length number distribution exponent $\varepsilon$
was also done by keeping $\alpha$ fixed at a value of 1.8.
As $\varepsilon$ increases, the shortest loops in the distribution
are weighted more strongly.
Thus, higher values of $\varepsilon$ produce more X-ray emission
from cooler and denser regions, and lower values of $\varepsilon$
produce emission from hotter and more tenuous regions.
For each star, $L_{\rm X}$ for the entire distribution of loops
decreases as $\varepsilon$ increases.
Figure 13 shows the resulting probabilities for a grid of
models computed for a range of $\varepsilon$ between 1 and 4
with an increment of 0.25.
The most successful model of these was for $\varepsilon = 2.75$,
with only a marginally better probability than for the standard
value of 2.5 (i.e., both of these models are near the peak of
the probability curve).

Finally, another set of models was computed to address the concerns
raised in Section 5 about a possible overestimation of the
deceleration experienced by accreting clumps.
The postshock temperatures $T_{\rm sh}$ shown in Table 2 are
generally lower than the typically reported range of 2--3 MK
for CTTS accretion shocks.
Thus, it is possible that the clump speed $v_c$ has been reduced
{\em too much} below the ballistic free-fall speed $v_{\rm ff}$
by the effects discussed in Section 3.
A set of accretion shock and coronal loop models were thus
computed under the assumption that no deceleration occurs
(i.e., that $v_{c} = v_{\rm ff}$).
The resulting X-ray luminosities from the accretion shock
regions were substantially larger than in the standard models,
with 11 out of the 28 stellar cases experiencing more than a
factor of 100 increase in $L_{\rm X}$ (with 6 of these 11 having
more than a factor of $10^4$ increase), and only 6 other cases
exhibiting no change in $L_{\rm X}$.
The photospheric Alfv\'{e}n wave amplitudes for these models
were significantly larger than in the standard models, and thus
the coronal-loop heating was larger.
Loop models with $\alpha = 1.75$ (which is the best-fitting model
with constant $\alpha$ in Figure 13) yielded an increase in
$L_{\rm X}$ of factors of order 2--6 for 10 of the 28 models and
either equal or slightly lower luminosities for 8 of the 28 cases.
Summing the shock and loop model luminosities and
performing the $\chi^{2}$ comparison with the observations,
as described above, yielded a substantially decreased
probability of agreement for these ``undecelerated'' models,
with $P = 9 \times 10^{-5}$ for the case of
$\alpha = 1.75$ and $\varepsilon = 2.5$.
This is a reduction in probability from the corresponding
standard model (see above) of about a factor of 300.
Similar results for other values of $\alpha$ lead to the
conclusion that {\em some degree of clump deceleration is
necessary} for optimal agreement with observed X-rays.

\section{Discussion and Conclusions}

The primary aim of this paper has been to explore and test a set
of physical processes that may be responsible for producing stellar
winds and coronal X-ray activity in accreting T Tauri stars.
A key new aspect of this work is the self-consistent generation
of plasma heating and wind acceleration (via MHD waves and
turbulence) for a database of well-observed CTTS in the
Taurus-Auriga region.
Scheurwater \& Kuijpers (1988) and Cranmer (2008) suggested that
inhomogeneous accretion can give rise to enhanced MHD wave
amplitudes across the stellar surface.
In this paper, the consequences of these turbulent motions
were studied further and applied to both wave-driven wind models
and footpoint-shaking models for coronal loops.
By and large, the predicted X-ray luminosities from a combination
of accretion shocks and hot loops agrees reasonably well with
existing observations.
The resulting mass loss rates for polar CTTS winds tend to be
lower than those derived from measurements.
This indicates that disk winds or X-winds (i.e., outflows not
rooted to the stellar surface) may be responsible for many of
those measurements.
However, in several of the modeled cases the stellar winds do
appear to contribute significantly to the measured mass loss rates.

The models presented above have shown that it is plausible to
assume that the accretion energy is sufficient to drive CTTS
stellar winds and coronal X-ray emission.
However, this cannot explain the even stronger X-ray emission
from WTTS, which are believed to not be accreting at all.
Even if the lower X-ray fluxes of CTTS were the result of
absorption in the accretion columns (e.g., Gregory et al.\  2007),
the relatively unattenuated X-rays of WTTS should not be
explainable by accretion-driven phenomena.
Here it was assumed that the accretion-driven component of
surface MHD turbulence dominates the intrinsic convection-driven
component.
Young, rapidly rotating stars are suspected of having more
vigorous convective motions, higher internal dynamo activity,
and more frequent magnetic flux emergence than older stars such
as the Sun (e.g., Montalb\'{a}n et al.\  2004; Ballot et al. 2007;
Holzwarth 2007; Brown et al.\  2008; K\"{a}pyl\"{a} et al.\  2009).
Rapid rotation can also give rise to complex field-line motions
that lead to propeller-driven outflows, collimated jets, or
``magnetic towers'' (Blandford \& Payne 1982; Lynden-Bell 2003;
Ustyugova et al.\  2006; Romanova et al.\  2009).
More accurate models must include these kinds of effects.

In this paper, an attempt was made to reduce the number of
freely adjustable parameters by as much as possible.
For example, although the accretion shock model (Section 5)
depends on many assumptions about the magnetic geometry and
the form of the radiative cooling function, it does not 
contain any truly ad~hoc free parameters.
The coronal loop model (Section 6) utilizes a state-of-the-art
parameterization for the heating that comes from MHD turbulent
dissipation, and the best value for the $\alpha$ exponent 
in that parameterization turns out to be one that has been
fit from a series of numerical simulations (e.g.,
equation [\ref{eq:alrapp}]).
However, all of the models presented in this paper do depend
on various assumptions that had to be made about the
clumpiness of the accreting gas, and about how these clumps
may be decelerated as they fall from the inner edge of the disk.
Future models of inhomogeneous accretion should focus on
improving our understanding of the dynamics of individual
clumps.

The use of a rotationally aligned dipole magnetic field in this
paper is a shortcoming that should be remedied in future models.
In fact, there has been much work done recently to create
three-dimensional models of CTTS magnetospheres (e.g.,
Koldoba et al.\  2002; Romanova et al.\  2003, 2004, 2008;
Gregory et al.\  2006, 2009; Long et al.\  2007, 2008).
Such models can naturally include the effects of X-ray absorption
by the accretion columns and the dynamics of disk winds or X-winds.
Still, many of these models utilize relatively simple approximations
for the conservation of internal energy.
Their realism and predictive power could be improved by including
the effects of MHD waves and turbulence as described above.
Thus, it is possible that the most useful results of this paper
may not be the specific model results, but instead the general
{\em methodology} for wave generation and coronal heating that
can be inserted into more advanced simulations.

In addition to improving the physical models of the stellar
plasmas, it will also be important to model the emergent
radiation more accurately.
For X-rays, the simulation of bandpass-integrated luminosities
(Section 7) is only the beginning of the story.
There is a great deal of diagnostic information in the X-ray
spectrum that can be used to constrain the properties of the
various plasma components.
Recent X-ray spectral analyses are beginning to reveal that
there may be {\em other} distinct emitting regions in addition
to the three studied in this paper (e.g., winds, accretion shocks,
and coronal loops).
One of these other classes includes transient phenomena such as
strong flares, eruptions, and other outcomes of large-scale
magnetic reconnection (e.g., Giardino et al.\  2006;
Kastner et al.\  2006; Aarnio et al.\  2009).
Another newly discovered region appears to be filled with
cooler and lower-density coronal plasma than is in the dominant
coronal loops (Brickhouse et al.\  2009).
These regions may be turbulent ``boundary layers'' that
surround the accretion spots and are magnetically connected
to both the shocked plasma and the loops (see also
Hujeirat \& Papaloizou 1998).

\hspace*{0.01in}

I gratefully acknowledge Nancy Brickhouse, Adriaan van Ballegooijen,
and Andrea Dupree for many valuable discussions, as well as
Sean Matt for helping me discover an error in an earlier version
of this paper.
This work was supported by the Sprague Fund of the Smithsonian
Institution Research Endowment, and by the National Aeronautics
and Space Administration (NASA) under grant {NNG\-04\-GE77G}
to the Smithsonian Astrophysical Observatory.

\end{document}